\documentclass[]{imag-ms-template}

\usepackage{booktabs}
\usepackage{graphicx}
\usepackage{subcaption}
\usepackage{multirow}
\usepackage{url}
\usepackage{makecell}
\title{Stronger Alignment between Brain Activity and LLM Embeddings during Code Writing compared to Prose Writing} 

\author{Zachary Karas$^{1\ast}$, Catie Chang$^{1}$, Kevin Leach$^{1}$, Yu Huang$^{1}$\\
{\small $^{1}$Department of Computer Science, Vanderbilt University,}\\
}

\begin{document} 

\maketitle


\begin{abstract}
Programming is a critical skill underlying modern software systems, yet the cognitive processes supporting code writing are only beginning to be understood,
limiting educational practices and developer tools.
At the same time, Large Language Models (LLMs) are increasingly used to assist programming.  
These models themselves are not well understood and can exhibit undesirable behavior like introducing security vulnerabilities.
Given evidence that some cognitive representations may be shared between LLMs and the brain, we seek to improve our understanding on both fronts by relating these two systems to one another.
We used Voxelwise Encoding Models (VEMs) to relate LLM embeddings to brain activity measured with functional Magnetic Resonance Imaging (fMRI) during naturalistic writing tasks. 
Using participants’ ($n=23$) keystrokes as prompts, we extracted LLM embeddings to predict voxelwise Blood Oxygen Level Dependent (BOLD) signal, quantifying alignment as the correlation between predicted and recorded signal. 
To assess whether this alignment is specific to programming or generalizes to other generative processes, we compared code writing to prose writing.
Alignment was strongest in the right frontal pole, and brain activity was significantly better predicted by LLM embeddings during code writing than prose writing ($p<0.001$, FDR-corrected).
Within
participants, the best-modeled voxel locations for code writing were 66\% consistent across LLM layers but varied substantially between participants (39\% similarity).
Our findings suggest stronger alignment between human and LLM representations during structured code generation, with implications for designing AI systems that predict code generation but support natural language tasks.
\end{abstract}

\section{Introduction}
Programming has a large influence on society today, where software plays integral roles in healthcare, infrastructure, finance, and entertainment~\citep{stoumpos2023digital, das2024exploring, kamuangu2024digital}.
The cognitive demands of writing, understanding, and debugging code have heightened as the size and complexity of software systems continue to grow, like that of Google, whose codebase was over 2 billion lines of code in 2016~\citep{potvin2016google}.
Supporting developers working on these systems often involves cognitive interventions, such as education to promote effective strategies for preventing mistakes and deconstructing complex computational problems into steps~\citep{haidry2017identifying, su2023systematic}. 
These interventions benefit from a stronger understanding of cognition, so researchers have used functional Magnetic Resonance Imaging (fMRI) in collaboration with Software Engineering (SE) researchers to investigate programming skills~\citep{siegmund2014understanding, krueger2020neurological, karas2021connecting}.
Those efforts have identified parietal and prefrontal regions that are involved in tasks like code writing and code reading~\citep{siegmund2014understanding, karas2021connecting, peitek2021program}, which has provided a foundation for interpreting programmer thinking and developing educational interventions~\citep{endres2021read}.
Further sharpening our understanding of the cognition underlying programming tasks has implications for education, where factors can be addressed that impede developers from decomposing a problem into steps, or implementing multi-step edits to defective code.

In parallel, many Artificial Intelligence (AI) tools for programming have recently been created to assist developers, promising to reduce their cognitive load during programming tasks~\citep{graaff2023let}.
However, these AI tools themselves are not well understood, and often exhibit unexpected or undesirable behavior~\citep{mastropaolo2025triumph}. 
For example, AI tools have been designed to generate code, find bugs, and test code~\citep{alenezi2025ai}, but they can also introduce security vulnerabilities~\citep{abdali2024securing}, pass software tests by changing the tests themselves~\citep{baker2025monitoring}, and fall prey to malicious prompt injection attacks~\citep{yu2023assessing}.
These AI tools are often inadequate and misaligned to developers' goals, but have the potential to address endemic problems in SE, such as translating outdated programming languages deeply ingrained in software systems to more modern, powerful, and secure languages~\citep{lei2023creating}.

As a result, efforts to support developers face serious challenges on two fronts: (1) we do not fully understand the cognitive mechanisms underlying code writing, which limits training and tool development, and (2) the AI tools designed to help developers are poorly understood and insufficient, which affects their adoption and trustworthiness.
Of particular interest, prior research has found similarities in how LLMs and the human brain represent semantic concepts~\citep{huth2012continuous}.
This link between LLMs and the human brain provides a potential avenue for relating these distant challenges by improving our understanding of both cognitive systems. 
Specifically, prior research has found that, similar to LLMs, objects and action categories are also represented in the human brain in a numerically continuous space, and that these representations can be mapped to one another using Voxelwise Encoding Models (VEMS)~\citep{huth2012continuous}.
While previous studies have involved passive language tasks like podcast listening~\citep{tang2023semantic}, to our knowledge, no previous study has used VEMs to model brain activity associated with code writing.
Thus, a central hypothesis of our work is that code writing processes in the brain also rely on a continuous space that can similarly be modeled using LLM embeddings.
In this study, we use VEMs to both improve our understanding of cognitive mechanisms underlying software skills, and provide insight into AI tools to improve their usability.

VEMs are a powerful method for studying cognition in lightly structured, naturalistic fMRI experiments.
Using this technique for movie watching, for instance, researchers can use the visual and auditory features of the film as a frame of reference for modeling the BOLD signal~\citep{ccukur2013attention}.
By relating these complex features to brain activity, this technique can provide detailed insights into cognitive processes ranging from low-level perception to high-level multisensory processing~\citep{dupre2025voxelwise, henderson2023texture}. 
To our knowledge, researchers have not used VEMs to study programming tasks, but this approach can potentially help to disentangle the complex cognitive mechanisms associated with programming.
In SE, studies have found using general linear models that activity in Broca's Area increases as programmers read more complex code~\citep{peitek2021program} and that a fronto-parietal logic network is enlisted for programming in novices~\citep{liu2025rapid}.
Researchers have also found using functional connectivity analysis that during code writing, there is increased connectivity between Broca's Area and the Number Form Area~\citep{karas2021connecting}. 
Those studies have uncovered relevant regions and networks for programming, but have struggled to disentangle the neural processes that give rise to programming.
While single-word stimuli can be sufficient for a language experiment~\citep{murphy2019neural}, or single operations for a mathematics experiment~\citep{arsalidou2018brain}, a single functional unit of code often consists of multiple lines that together perform one unified role.
Thus, studying cognitive processes with respect to the interconnected structure of code could benefit from using VEMs, which could preserve this complexity in contextualizing the brain activity. 

The process for computing VEMs typically involves creating a feature representation of the task stimuli, then using ridge regression to train a linear mapping that translates between this \textit{feature vector} and the BOLD signal~\citep{huth2016natural}. 
Once this VEM is trained, new feature vectors can be used to predict corresponding brain activity.
In a podcast-listening study, for example, a trained VEM could accurately predict a participant's brain activity in specific regions from a feature vector of the transcript~\citep{tang2023semantic}.
Researchers have recently been creating these vector representations using LLMs; for instance, Tang et al.  prompted GPT-2 with a podcast transcript, then extracted vector embeddings from the model~\citep{tang2023semantic}.
This approach leverages the ``cognition'' of LLMs, which learn to represent words and concepts numerically, where similar words cluster together in high dimensional space based on nuances in their semantic meanings (i.e., carrot, onion)~\citep{stankevivcius2024extracting}.

In this paper, we use VEMs to investigate code writing in human programmers ($n=23$).
We leveraged participants' \textit{keystrokes} during naturalistic writing tasks from a previously collected dataset~\citep{krueger2020neurological}, using their written responses for each moment in time as prompts for LLMs. 
We then extracted the models' internal vector embeddings to serve as a rich abstraction of participants' written responses for ridge regression, which learned a linear mapping between these embeddings and participants' voxelwise BOLD signal. 
To contextualize our findings about predicting brain activity from the \textit{code} writing, we also used VEMs to predict brain activity from \textit{prose} writing in the same experiment, where participants answered long response questions in English; differences here can demonstrate whether alignment between LLMs and brain activity may be task-dependent.
Furthermore, with the rapid development of state-of-the-art LLMs, researchers can choose from an increasingly diverse set of models to train VEMs.
We therefore tested the performance of six different LLMs in predicting BOLD activity for both code and prose writing, as well as eight different layers within each LLM. 

A large body of research has investigated how linguistic features are encoded and processed by different LLM layers~\citep{lopez2025linguistic}, but has largely focused on BERT models, which are outperformed by LLMs today.
Early layers in BERT models have been found to represent low-level information while later layers have been found to represent high-level semantic concepts or features like metaphors~\citep{vaidya2022self, tenney2019bert, lopez2025linguistic}.
This selectivity is also reflected in how these layers can predict BOLD signal in regions of the human brain~\citep{vaidya2022self}. 
For instance, Vaidya et al. found evidence suggesting that activity in the auditory cortex was better predicted by early LLM layers, while intermediate layers were better able to predict activity in the middle temporal gyri and Visual Word Form Area~\citep{vaidya2022self}.
To test the selectivity of current LLM layers for modeling different cognitive processes, we analyzed whether certain layers were better able to predict activity in different brain regions.

As such, we organize our lines of inquiry into the following research questions:
\begin{enumerate}[noitemsep]
    \bfseries
    \item How does voxelwise encoding model performance differ between code and prose writing?
    \item How do behavioral and experimental factors affect the alignment between LLM embeddings and BOLD signal during code and prose writing?
    \item Do embeddings from different LLM layers better predict activity in different brain regions?
\end{enumerate}

We found that VEMs effectively modeled participants' voxelwise BOLD signal during writing tasks, but were significantly better
for code writing compared to prose writing ($p<0.001$, FDR-corrected).  
We report our highest performing parameter combinations for the models we considered.
DeepSeek 6B and StarCoder 3B performed best for code writing, while StarCoder 7B performed best for prose writing.
Surprisingly, for both code and prose writing, we found a high level of consistency across LLM layers in the brain regions where top-modeled voxels were located.
We observed this consistency \textit{within} participants, where the top-modeled regions for code writing were 66\% consistent across model layers for a given participant; however, the top-modeled regions were only 39\% similar \textit{between} participants.
These findings were surprising in the context of prior literature, 
and also suggest that there may be a unique ``fingerprint'' for how a participant's brain activity during code and prose writing can be modeled by LLM embeddings, which appears to be consistent across the model layers.

Overall, the contributions of this study include the following:
\begin{itemize}[noitemsep]
    \item Introduces the use of VEMs to model brain activity associated with naturalistic code and prose writing.
    \item Provides a systematic analysis between different study parameters, totaling 55,200 different VEMs trained.
    \item Demonstrates that brain activity in the right frontal pole (i.e., delayed intentions, information integration) can be grounded in participants' writing behavior.
    \item Provides insights for AI explainability suggesting that information processing is distributed across modern LLM layers.
    \item Discusses implications for AI tools suggesting LLMs can help \textit{predict} code and identify when programmers stray from expected patterns, but \textit{collaborate} on prose support the user.
\end{itemize}

\section{Methods}\label{sec:methods}
\begin{figure}[htbp!]
    \includegraphics[width=\linewidth]{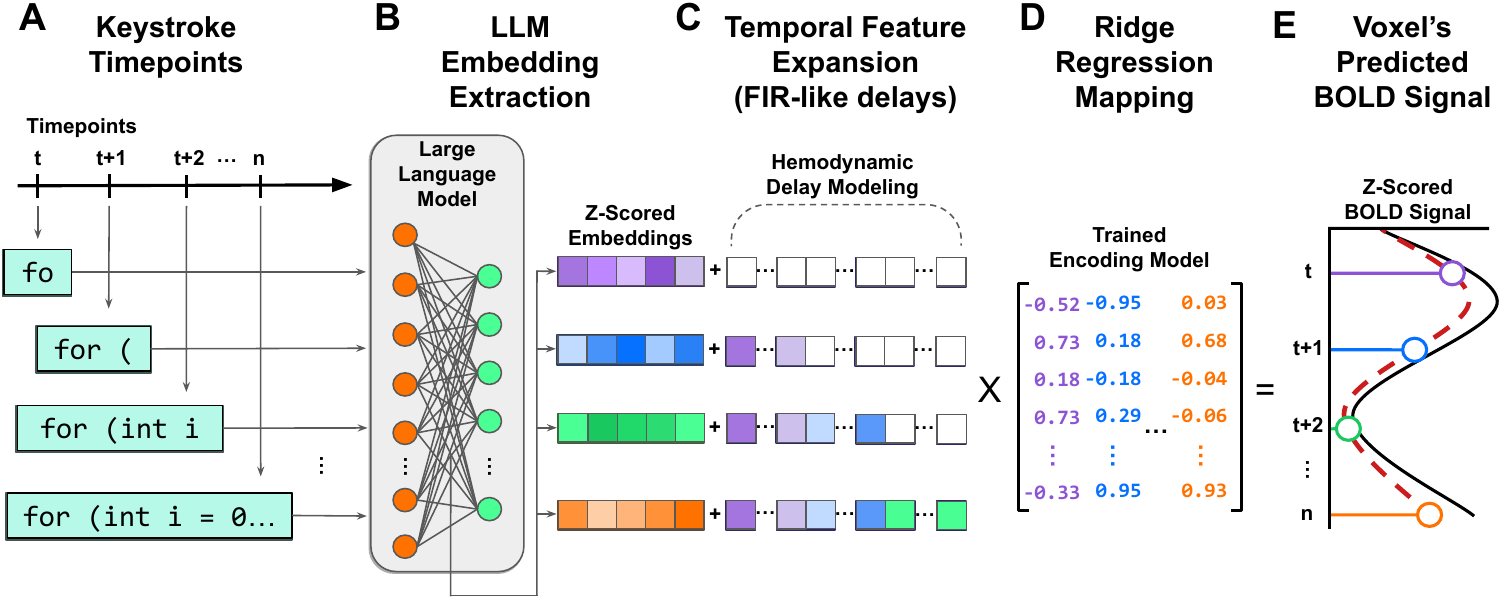}
    \caption{Conceptual diagram of our methodology. (A) At each fMRI timepoint, we used participants' keystrokes as a prompt into an LLM, then (B) extracted internal embeddings that served as a numerical abstraction of the keystrokes. We included additional context, described below, and then (C) concatenated delayed copies of the feature vectors to model the hemodynamic delay (with zero-padding). We then (D) used ridge regression to train a voxelwise encoding model to map these feature vectors to each voxel's BOLD signal, which enabled us to (E) predict BOLD signal from keystroke embeddings for a held-out test set.}
    \label{fig:methods}
\end{figure}

In this section, we describe the fMRI dataset and analysis methods used in this study, including details about the original study, the steps for preprocessing the fMRI data and keystrokes, and the approach for computing VEMs.
Conceptually, the framework for predicting BOLD signal from LLM embeddings is similar to the Finite Impulse Response (FIR) model of the hemodynamic response, where features of the stimulus are combined and weighted to model the BOLD signal value at a given timepoint~\citep{henson2007convolution}.
Instead of designing these features by hand, the features here are LLM embeddings, which are numerical vectors that serve as a richly encoded representation of the task stimulus, but are still combined and weighted to predict the BOLD signal for a given timepoint~\citep{dupre2025voxelwise}.
Here, we prompted LLMs with participants' keystrokes at each timepoint, and extracted embeddings to obtain a vector representation for that timepoint of what the participant wrote. This process is depicted in Figure~\ref{fig:methods}.

\subsection{Dataset Preparation}
To study the alignment between LLM embeddings and human brain activity during code and prose writing, we used a previously recorded fMRI dataset for which undergraduate and graduate computer science students ($n=25$) from the University of Michigan wrote code and prose in response to different questions~\citep{krueger2020neurological}\footnote{More details related to the experimental procedure are included in the original study, with replication materials available here: \url{https://web.eecs.umich.edu/~weimerw/fmri.html}}.
Data was excluded from two participants who completed only one of the two tasks since a primary objective of our study was to compare performance of modeling brain activity during code and prose writing.
This yielded data from 23 participants for our study.
Select demographic information about the participants in that IRB-approved (HUM00138634) study are shown in Table~\ref{tab:demo}.
Informed consent was obtained from all participants for being included in the study.
Broadly, the \textbf{Code} condition from that study asked participants to write simple functions, statements, or conditional statements in the C++ programming language while the \textbf{Prose} condition asked participants to describe their opinions on matters ranging from working on a team to level of comfort with change.
An example coding question asked ``\textit{Write the definition of a function} \texttt{isPositive}\textit{, that receives an integer parameter and returns true if the parameter is positive, and false otherwise.}''
A prose question stated ``\textit{Some people prefer to spend time with one or two close friends. Others choose to spend time with a large number of friends. Compare the advantages of each choice.}''
Participants were given 60 seconds to answer each question, of which there were nine for both the Code and Prose conditions.
The code and prose questions were separated into their own blocks that each took 15 minutes to complete with jittered rest in between each question.

\begin{table}[htpb!]
\small
    \centering
    \caption{Demographics for our sample of participants ($n=23$).}
    \label{tab:demo}
    \begin{tabular}{c|c|c|c|c} 
    \hline
    \textbf{Avg. Age} & \textbf{Women} & \textbf{Native English Speakers} & \textbf{Coding Experience (years)} & \textbf{Avg. GPA} \\\hline
        21.05 & 6 & 20 & 2.5 & 3.44 \\
    \bottomrule
    \end{tabular}
    \begin{tablenotes}
        \small
        \centering
        \item GPA: Grade Point Average
    \end{tablenotes}
\end{table}

\textbf{Keyboard}
A 101-key QWERTY USB keyboard was adapted to work in the Magnetic Resonance environment. 
More details are provided in the original study~\citep{krueger2020neurological}, but the metal components and control logic were removed from the keyboard and placed outside the scanning room. 
Moving metal pieces in the keyboard were replaced by 3D-printed plastic, and each key was attached to its own shielded wire to limit electromagnetic interference.
The wires were then fed through the control panel, where a custom-built device read the signal from each key press and output a standard USB signal. 
Placing this control logic outside of the scanning room eliminated electromagnetic interference during the scans, and enabled live text editing for participants.
The keyboard occasionally exhibited erroneous behavior where certain characters would be typed repeatedly without any participant input. 
We identified and removed five timepoints across three participants associated with erroneous keystrokes.

\textbf{Scanning Parameters} Scans were collected on a 3T MR750 GE Scanner with a 32-channel head coil. 
High-resolution anatomical scans ($T_1$-weighted spoiled gradient recalled sequence, 208 slices, $1mm$ thickness) were collected first with $TR$ = $2300.80ms$, $TE$ = $24ms$, $TI$ = $975ms$, and $FA$ = $8^\circ$. 
An estimate of the magnetic field homogeneity within the scanner was collected using a spin-echo fieldmap with a $TR$ = $7400ms$, $TE$ = $80ms$, and slice thickness of $2.4mm$. The functional data ($T_{2}^{*}$-weighted multiband echo planar imaging sequence, 60 slices, $2.4mm$ thickness) was collected with a $TR$ = $800ms$, $TE$ = $30ms$, $FA$ = $52^\circ$, and an acceleration factor of 6. Including slice thickness, the isotropic voxel dimensions were [$2.4mm\times{2.4mm}\times{2.4mm}$]. 

\subsection{Preprocessing}\label{preprocess}
In this study, we used participants' keystrokes to contextualize the functional data, so here we describe our methods for preparing both the fMRI data and keystrokes for training VEMs using ridge regression.

\textbf{fMRI Data.}
We implemented a standard preprocessing pipeline for the fMRI data using FSL~\citep{jenkinson2012fsl}, ANTs~\citep{tustison_antsx_2021}, and custom scripts written in bash and MATLAB. 
The first 12 volumes were removed for T1 equilibration. 
The scans were then slice-timing corrected and unwarped. 
We extracted the brain from the anatomical images, then performed affine and non-linear registration of the anatomical images to the MNI152 template as a target.
After performing brain extraction on the functional data, we applied motion correction and recorded motion parameters along 6 axes~\citep{jenkinson2012fsl, tange2018gnu}. 
We then applied the affine and non-linear transformations to the functional data, and did not apply gaussian smoothing.
Independent Component Analysis (ICA) was performed to remove additional artifacts.
The quality of physiological data recorded from the scanner was low, so the ICA step also served to correct for physiological noise, to the extent that physiological effects are represented in spatial ICA components. 
We decomposed each participant's functional scan into 60 components~\citep{jenkinson2012fsl}, then the first author (4 years of experience assessing BOLD signal quality) hand-classified each component as signal or noise for 16 participants~\citep{griffanti2017hand}.
These labels were then used to train a classifier specific to this data, which we used to identify and remove noisy components from all participants' functional data (threshold of 20).
Next, we regressed out the 6 motion parameters, their first derivatives, and a linear trend from each voxel's timecourse. 
Finally, we z-scored the functional data along the temporal axis for each participant, which increases stability for machine learning~\citep{huth2016natural}.

\textbf{Keystrokes.}
Our goal for preprocessing the keystrokes was to align the keys participants pressed with the fMRI volumes that were recorded at the same time. 
Participants answered multiple questions, so the aligned keystrokes needed to be specific to each question, and we needed to process special characters like arrow keys and backspaces to obtain accurate representations of participants' answers that would serve as prompts for LLMs.
The keystrokes were recorded as raw ASCII characters with associated timestamps, where each key is represented as a number (i.e., A: 65, a: 97, Space: 32). 
We first aligned the keystrokes to corresponding volumes in the fMRI data using recorded onset times, individually for each question. 
With a TR of 800 milliseconds, we identified the keys that were pressed within start and end times for each volume, then translated each ASCII code into its corresponding character. 

Once the keystrokes were properly aligned to each volume, we processed the special characters like shift, enter, backspace, and arrow keys to obtain an accurate representation of what participants wrote. 
Line numbers and cursor position were tracked as text was added or deleted, and as participants pressed the enter key or arrow keys, which ensured that text would be added or deleted in the correct location. 
This applied to text that would be deleted from a preceding line if participants deleted all the text on the current line. 
Multiple shift keys from participants holding the key down were treated as a single instance of shift, and were used to translate the succeeding key into either a capital letter or the corresponding symbol (i.e., $1\rightarrow~!$). 
Following this process yielded volume-aligned keystrokes that accurately reflected what participants typed.

\textbf{LLM Prompts from Keystrokes.}
In the previous step, we aligned keystrokes participants wrote with each volume.
We next added additional context to improve interpretability by the LLM ~\citep{brown2020language}.
Specifically, for a given timepoint we included the current question text as well as the participant's answer up to that point.
For instance, the characters ``\texttt{own}'' that a participant wrote during an 800ms time window in the middle of a scan may not convey important contextual information to an LLM.
After including the question text and the participant's previous keystrokes, the characters would instead be the following: ``Write a function that counts down from ten to one: \texttt{void countD[own]}.''
Furthermore, participants in the middle of writing a word likely knew what they were going to type next (i.e., typing ``\texttt{eat}'' in ``\texttt{greatest}''), so we also included the future keystrokes participants typed for a tuneable number of succeeding timepoints.
More broadly, including future keystrokes can perhaps model participants' planning behaviors, albeit naively, so we modified the amount of future timepoints that we included to test its influence on modeling performance.
Specifically, we constructed copies of the keystrokes that ``looked ahead'' by a log-spaced 0, 1, 3, 5, and 10 volumes to consider a broad range of possibilities for planning behavior.

To delineate the text from the past, the future, and the current timepoint, we adopted a method used to train LLMs on text data called \textit{fill-in-the-middle}, where the characters are denoted with tags to represent the Prefix ($<$PRE$>$), the Suffix ($<$SUF$>$), and the Middle ($<$MID$>$), respectively~\citep{bavarian2022efficient}.
Since LLMs are trained to predict the next token in a sequence, the middle token is put last for this training technique so the model can learn to predict an intermediate token.
As such, the finalized keystrokes for a given timepoint to be used as an LLM prompt would be formatted as follows: ``$<$PRE$>$Write a function that counts down from ten to one: \texttt{void countD}$<$SUF$>$() \{$<$MID$>$own''.
We used this format to provide context to the LLMs and improve the interpretability of participants' keystrokes since it follows data formatting techniques for training LLMs~\citep{bavarian2022efficient}.

\textbf{LLM Embeddings.}
After formatting participants' keystrokes for each timepoint, we used these keystrokes as prompts into LLMs, then extracted the embeddings to use as feature vectors for contextualizing the voxelwise BOLD signal.
There are many state of the art models and different sizes among the models, so we considered two sizes of three different LLMs to systematically compare six total models and understand trends between the models and their sizes.
Furthermore, the LLMs we considered are comprised of up to 32 layers from which we could extract the embeddings.
Primarily in BERT models, researchers have found evidence that early layers process low-level lexical information, while later layers process high-level semantic concepts~\citep{vaidya2022self, toneva2019interpreting, tenney2019bert}, so we considered eight layers evenly spaced throughout each of our chosen models to test whether the embeddings from these layers are better able to model the BOLD signal in different brain regions.
The models we considered, their sizes, and the layers we considered from each are included in Table~\ref{tab:models}.
We did not include GPT models in our analyses because we only used state of the art models that are open source, allowing users access to model parameters.

\begin{table}[htbp]
    \centering
    \begin{tabular}{l|l}
        \hline
        \textbf{Model} & \textbf{Layer Numbers Considered} \\\hline
            CodeGemma  2B & \phantom{0}0, \phantom{0}2, \phantom{0}5, \phantom{0}7, 10, 12, 15, 18 \\
            CodeGemma 7B  & \phantom{0}0, \phantom{0}4, \phantom{0}8, 12, 16, 20, 24, 28 \\
            DeepSeek 2B   & \phantom{0}0, \phantom{0}3, \phantom{0}6, 10, 13, 17, 20, 24 \\
            DeepSeek 6B   & \phantom{0}0, \phantom{0}4, \phantom{0}9, 13, 18, 22, 27, 32 \\
            StarCoder2 3B & \phantom{0}0, \phantom{0}4, \phantom{0}8, 12, 17, 21, 25, 30 \\
            StarCoder2 7B & \phantom{0}0, \phantom{0}4, \phantom{0}9, 13, 18, 22, 27, 32 \\
         \bottomrule
    \end{tabular}
    \caption{The LLMs we used for this study, as well as the layers therein from which we extracted embeddings. These chosen layers were evenly spaced throughout the different models.}
    \label{tab:models}
\end{table}

The architecture of LLMs is complex, but they are comprised of multiple repeated blocks that generally contain an attention component and a Multi-Layer Perceptron (MLP) component~\citep{vaswani2017attention}.
We refer to these blocks as layers throughout this paper. 
In extracting embeddings, we followed previous research by using the MLP weights~\citep{tang2023semantic}. 
The MLP weights are formatted as a large two-dimensional vector, and related studies have either taken an average across the columns~\citep{srikant2022convergent}, or extracted the final row~\citep{tang2023semantic}, which behaves as a summary token by attending to each other token in the sequence~\citep{lewis2020bart}.
We chose the latter and extracted the final row of MLP weights from a given layer to serve as an LLM representation for our analyses.

These vectors are different lengths depending on the models (i.e., 3072, 4096 elements), so we applied Principal Component Analysis (PCA) to reduce their dimensionality and limit computational costs of ridge regression~\citep{aw2023instruction}.
We applied PCA such that the final vectors retained 99\% of the variance of the original vectors, which in itself uncovered interesting differences between code and prose writing.
The original vectors from the LLMs contained between 2048 to 4608 elements, depending on the model, and the average embedding length after applying PCA was 376.2 for Code, and 499.3 for Prose.
We found that this difference was statistically significant based on a paired t-test ($t=10.694$, $p<0.001$, $d=2.230$).
This finding suggests that the semantic richness of prose may be more difficult to reduce than the structured nature of code, which foreshadows our findings below in Section~\ref{results}.

To prepare these LLM embeddings for ridge regression, we included a nuisance regressor to model the motor processes of typing keys, which was simply the number of keys that a participant pressed as a volume was recorded. 
In computing these tallies, we counted multiple occurrences of Shift and Control as one key press, since these are held down during typing.
Lastly, we z-scored the vectors to improve stability in this machine learning context, as we did above for the fMRI data~\citep{huth2016natural}.
These vectors served as our feature vectors which we related to BOLD signal using ridge regression.

\subsection{Voxelwise Encoding Models}
Following prior research, we used ridge regression to learn a linear transformation between LLM embeddings and BOLD signal~\citep{dupre2025voxelwise}.
As discussed above, our LLM embeddings served as a richly encoded feature vector of the keystrokes participants typed at each timepoint.
Ridge regression then learns to predict, for each voxel individually, the value of the BOLD signal at each timepoint as a weighted linear combination of elements in the corresponding feature vectors~\citep{henson2007convolution}.
Importantly, ridge regression includes an \textit{L2} regularization term that penalizes large weights and prevents overfitting~\citep{huth2016natural}.
We account for the hemodynamic delay by concatenating shifted copies of the feature vector at successive timepoints~\citep{dupre2025voxelwise}, thus enabling the feature vector to influence the prediction of the BOLD signal at later timepoints~\citep{dupre2025voxelwise}.

We followed the methods of Huth et al. in constructing our ridge regression pipeline~\citep{huth2016natural}. 
That study uses a \textit{TR} of 2 seconds and includes four delayed copies of the feature vector,  modeling eight seconds total.
However, in other studies that use FIR modeling of the hemodynamic response, the number of lagged copies may be chosen to span the duration of the hemodynamic response function ($\approx$20 seconds)~\citep{yacubian2006dissociable, kay2008modeling}. In our analyses, we varied the number of delayed copies of the feature vectors to understand the tradeoffs between modeling flexibility and performance.
Specifically, we constructed feature vectors that included 0, 4, 10, 16, and 20 delayed copies of the feature vectors. 
This range includes the four delayed copies considered in previous research~\citep{tang2023semantic, huth2016natural}, as well as up to 16 seconds, which captures more of the hemodynamic response function~\citep{yacubian2006dissociable, kay2008modeling}.
We also conducted a systematic comparison across key modeling parameters (i.e., model size and layer) to obtain a holistic understanding of how these parameters influence the performance of the VEMs.
If we consider a single ridge regression model as the model trained between one set of LLM embeddings and one participant's fMRI data on one task (i.e., code or prose), here we compute $55,200$ ridge regression models ($23~participants\times2~tasks\times6~models\times8~layers\times5~delay~values\times5~look$-$ahead~values$).

To focus on relevant cortical regions and reduce computational costs, we computed ridge regression models only for the voxels with located within parcels of the Schaefer Atlas, which totaled 131,065 voxels.
In line with prior work, we used 90\% of the data for training the ridge regression models and reserved the remaining 10\% for testing~\citep{huth2016natural}. 
There were 746 volumes in each of the fMRI scans (code and prose were separate scans), and we excluded the jittered volumes during the rest periods, which ranged between 51 and 63 volumes in total.
Thus, each scan contributed between 615 and 626 volumes and associated keystroke embeddings for training, and between 68 and 69 volumes for testing\footnote{Scans for both Code and Prose were terminated early for Participant 109 due to technical difficulties, yielding 410 volumes (369 training, 41 testing) for Code and 457 volumes (402 training, 55 testing) for Prose.}.
Ridge regression includes a hyperparameter $\alpha$ that determines the strength of the $L2$ penalty term.
We determined the value of $\alpha$ for each ridge regression model using five-fold cross-validation within the training set, evaluating 12 possible $\alpha$ values between 10 and 10,000\footnote{$\alpha$ values of 10, 18.738, 35.112, 65.793, 123.285, 231.013, 432.876, 811.131, 1519.911, 2848.036, 5336.699, 10,000}.
After training, we evaluated each VEM on the corresponding held-out test set. The BOLD signal predicted from keystroke embeddings was computed by taking the dot product between the embeddings and the trained model weights, and performance was quantified for each voxel as the Pearson Correlation Coefficient between the predicted and recorded BOLD signal~\citep{dupre2025voxelwise}.
The correlation coefficients in each voxel serve as the basis for our downstream analyses.
For each participant, we also identified  the top 10K voxels with the highest correlation coefficients to understand where the model performs best, following prior voxelwise encoding studies ~\citep{huth2016natural, tang2023semantic}.

\section{Results}\label{results}
In this section, we present our findings from computing VEMs based on participants' keystrokes during code and prose writing tasks. 

\subsection{Modeling Code Writing Versus Prose Writing}\label{rq:parameters}
\begin{figure*}[htbp]
    \centering
    \includegraphics[width=\textwidth]{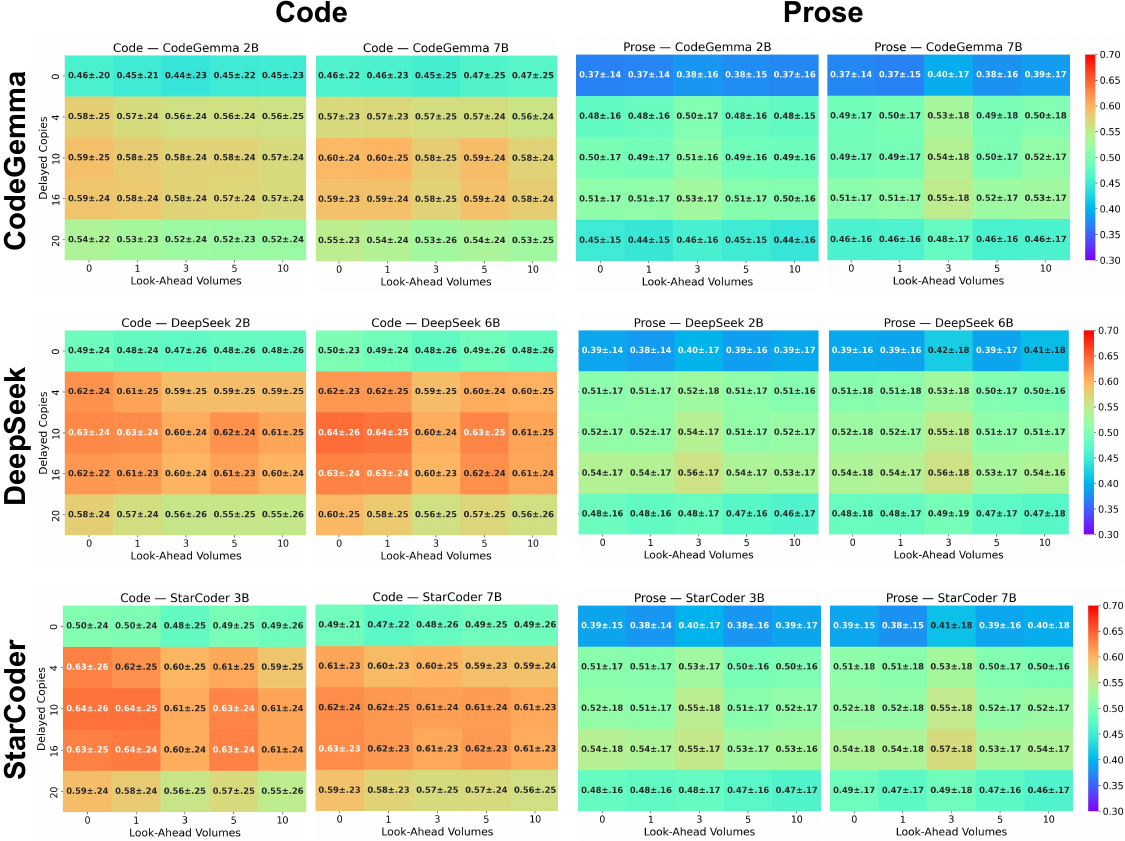}
    \caption{Modeling performance for Code and Prose for the six different models we used and each parameter combination we considered. We modulated the number of delayed copies of the feature vector we used to model the hemodynamic delay [0, 4, 10, 16, 20], as well as the number of volumes by which we ``looked ahead'' in formatting participants' keystrokes [0, 1, 3, 5, 10], which we use as a proxy
    to study planning behavior. Modeling performance was calculated by first z-transforming each participant's correlation coefficients measured between the predicted and recorded BOLD signal, filtering these coefficients to the top 10K, then computing the average for a given parameter combination.}
    \label{fig:heatmap}
\end{figure*}

For this line of questioning, we analyzed modeling performance for the various parameter combinations to gain a broad understanding of how well LLM embeddings could be used to predict voxelwise BOLD signal during code and prose writing.
To this end, we calculated the average of all participant's top 10K correlation coefficients for each model and each configuration of parameters (i.e., number of delayed copies, look ahead volumes), separately for Code and Prose. 
Correlation coefficients for Code and Prose and the different model families we considered can be seen in Figure~\ref{fig:heatmap}.
For each of these parameter configurations, we pooled the correlation coefficients together across the different LLM layers, and investigated performance related to individual layers below.
We see that correlations are higher for each configuration of parameters for Code compared to Prose.

\begin{table}[htbp]
    \small
    \centering
    \caption{Average modeling performance using embeddings from each LLM to predict BOLD signal during code and prose writing. Modeling performance was quantified as z-transformed correlation coefficients, filtered to the top 10K voxels for a given participant. Averages for one model were computed across layers, participants, and parameter configurations. Paired t-tests confirmed that these differences between code and prose modeling are highly significant.}
    \label{tab:paired_t_tests}
    \resizebox{\textwidth}{!}{
    \begin{tabular}{l|r|r|r|r|r}
        \hline
        \textbf{Model} & \textbf{Code Mean$\pm$St. Dev.} & \textbf{Prose Mean$\pm$St. Dev.} & \textbf{$t$-value} & \textbf{$p$-value} & \textbf{Effect Size ($d$)}\\\hline
            CodeGemma 2B & $0.540\pm0.050$ & $0.463\pm0.051$ & $31.133$ & $6.458e$-$21$ & $6.227$\\
            CodeGemma 7B & $0.547\pm0.048$ & $0.476\pm0.054$ & $17.431$ & $3.954e$-$15$ & $3.486$\\
            DeepSeek 2B & $0.574\pm0.053$ & $0.489\pm0.055$ & $25.300$ & $8.139e$-$19$ & $5.060$\\
            DeepSeek 6B & $0.582\pm0.053$ & $0.491\pm0.052$ & $21.463$ & $3.589e$-$17$ & $4.293$\\
            StarCoder 3B & $0.584\pm0.053$ & $0.486\pm0.055$ & $23.783$ & $3.398e$-$18$ & $4.757$\\
            StarCoder 7B & $0.577\pm0.051$ & $0.490\pm0.055$ & $28.252$ & $6.253e$-$20$ & $5.650$\\
         \bottomrule
    \end{tabular}
    }
\end{table}

These results are highly statistically significant as well, as demonstrated by paired t-tests for the average correlation coefficients for each model at each parameter configuration. 
Average correlation coefficients and results for paired t-tests are shown in Table~\ref{tab:paired_t_tests}.
Our findings suggest that VEMs using LLM embeddings are better able to predict brain activity associated with code writing than with prose writing.
Code writing is less expressive than prose writing, so these findings match our intuition that brain activity during code writing may better align with LLM embeddings than that during prose writing.
Furthermore, code has been found to be more predictable than the latter due to its highly structured syntax~\citep{hindle2016naturalness}.
Thus, in contrast to prose writing, it is possible that the more deterministic nature of code writing may culminate in representations that are more similar in the human brain and LLMs.

More qualitatively, we also see the optimal parameter configurations for the combinations we considered.
There may be a global optimum of parameters that we did not consider, but we consistently obtain higher correlation coefficients across the model families for Code by adding 10 and 16 delayed copies of the feature vectors, and by looking ahead by one or no volumes. 
The model families are notably consistent for Prose, where we obtain the highest correlation coefficients for each by including 16 delayed copies of the feature vectors and looking ahead by three volumes.
These findings demonstrate nuances between writing code and prose, where modeling prose writing may benefit more from looking ahead at participants' future keystrokes, perhaps because the increased context is more informative in Prose for the trajectory of participants' answers.
Consistent between Code and Prose, we see that adding no delayed copies of the feature vectors or 20 copies of the feature vectors yields lower correlation coefficients between the recorded and predicted signal. 
This may relate to practical considerations for modeling the BOLD signal, where no delayed copies of the feature vectors cannot accurately model the sustained hemodynamic response, while 20 delays may allow for too much flexibility~\citep{goutte2000modeling}.

\begin{figure*}[htbp!]
    \centering
    \begin{subfigure}[t]{0.48\textwidth}
        \centering  
        \includegraphics[width=\textwidth]{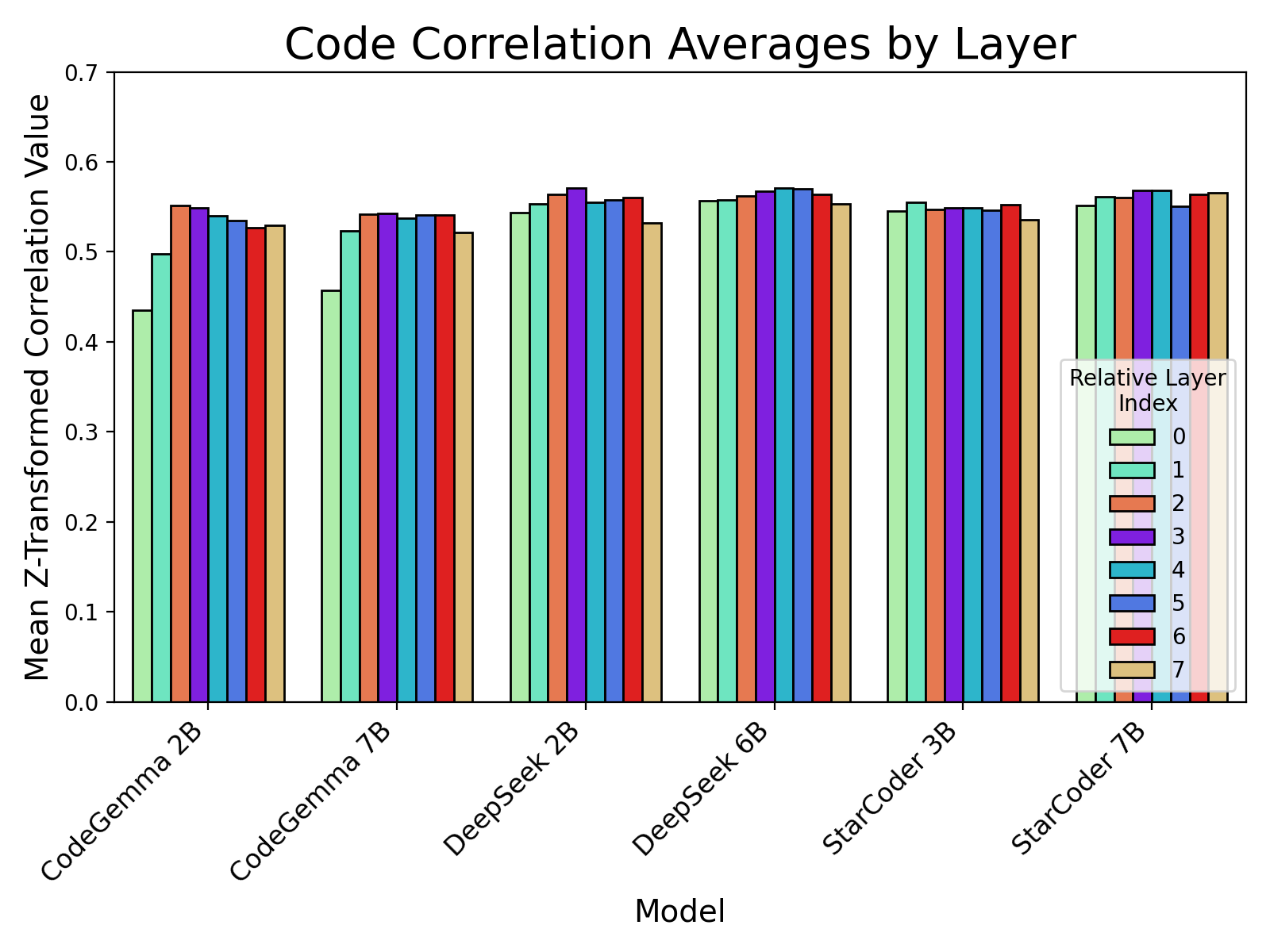}
        \caption{}
        \label{}
    \end{subfigure}
    \hfill
    \begin{subfigure}[t]{0.48\textwidth}
        \centering
        \includegraphics[width=\textwidth]{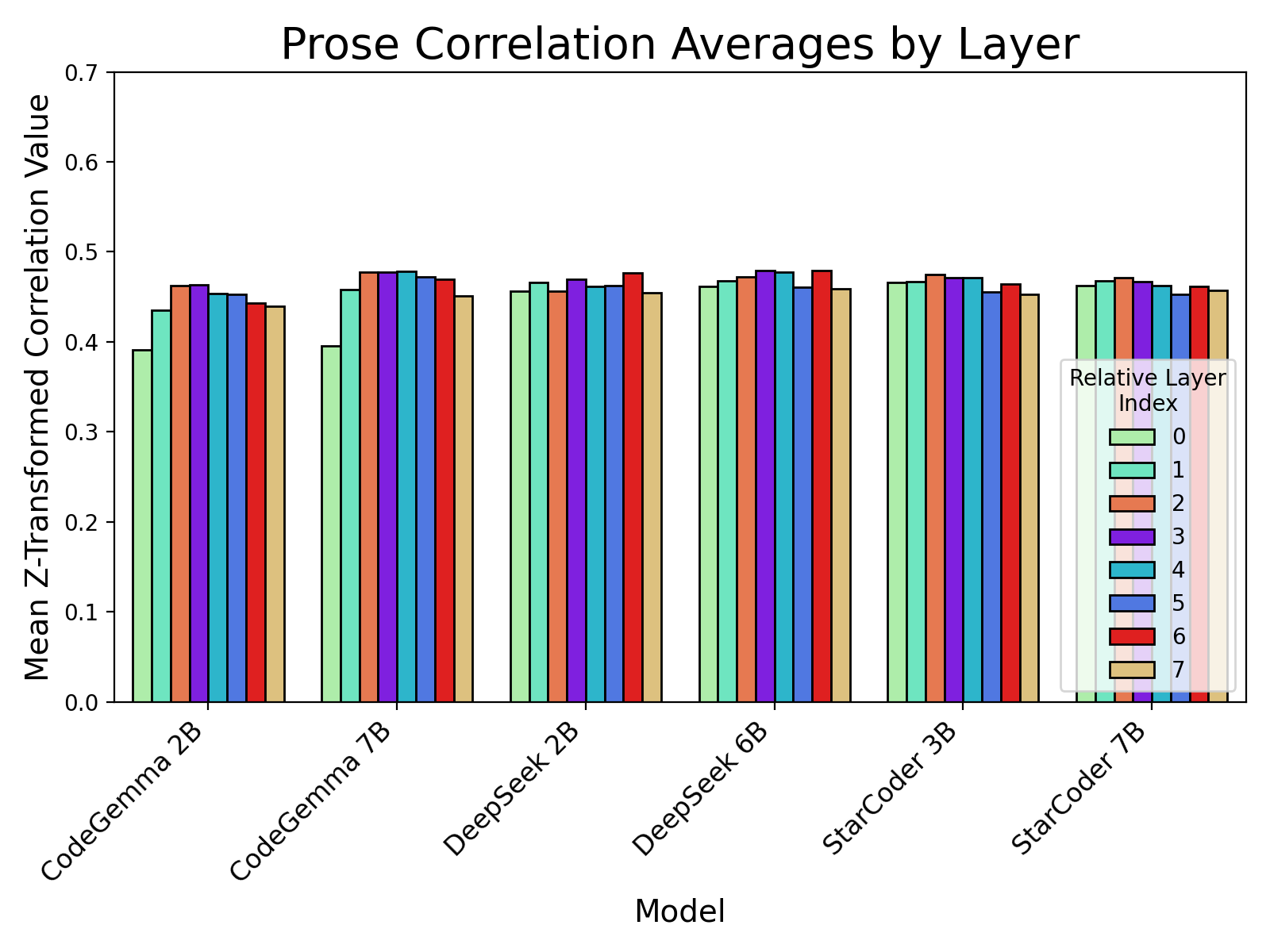}
        \caption{}
        \label{}
    \end{subfigure}
    \caption{Average modeling performance across the eight layers spaced throughout each of the six models we used, separated between (a) Code and (b) Prose.}
    \label{fig:layers_by_model}
    
\end{figure*}

These cumulative analyses demonstrate trends in how well different parameter configurations can be used to model voxelwise BOLD signal, and we next examined whether any LLM layers perform best for VEMs during code and prose writing.
Previous research has reported evidence that early and late linguistic processes are better represented in different LLM layers~\citep{vaidya2022self, lopez2025linguistic}, which is also reflected in how these layers can predict BOLD signal in regions of the human brain~\citep{vaidya2022self}. 
In this study on alignment between LLM embeddings and brain activity during code and prose writing, we investigated whether there are differences in brain regions that are best modeled by various LLM layers.
Based on this previous research, we hypothesized that early layers may perform better in modeling activity in low-level perceptual areas such as the motor cortex in our case, whereas middle and later layers may perform better in modeling activity in high-order frontal regions, or parietal regions associated with spatio-temporal processing~\citep{vaidya2022self, huang2019distilling, karas2021connecting}.

We conduct more thorough investigations into nuances between layers in Section~\ref{rq:layers}, but first analyzed whether there were overall differences between the various LLM layers in their modeling performance of voxelwise BOLD signal. 
For this purpose, we computed the average for each of the eight layers from the z-transformed correlation coefficients for the top 10K voxels from each participant, each of the six models, and each parameter combination. 
The results are depicted in Figure~\ref{fig:layers_by_model}, but performance is notably consistent across the layers for both Code and Prose.
This observation is supported by statistical evidence as well, where we conducted ANOVA tests for each model to see if there was a group-level difference between the various layers' correlation coefficients, but no tests reached statistical significance. 
Specifically, for Code we find no statistically significant group-level differences for 
CodeGemma 2B ($F=0.654$, $p=0.710$),
CodeGemma 7B ($F=0.538$, $p=0.805$),
DeepSeek 2B ($F=0.032$, $p=1.000$),
DeepSeek 6B ($F=0.018$, $p=1.000$),
StarCoder 3B ($F=0.008$, $p=1.000$), or
StarCoder 7B ($F=0.018$, $p=1.000$).
Similarly, for Prose we find no statistically significant group-level differences for 
CodeGemma 2B ($F=1.201$, $p=0.305$),
CodeGemma 7B ($F=0.753$, $p=0.628$),
DeepSeek 2B ($F=0.074$, $p=0.999$),
DeepSeek 6B ($F=0.043$, $p=1.000$),
StarCoder 3B ($F=0.134$, $p=0.996$), or
StarCoder 7B ($F=0.083$, $p=0.999$).
After correcting for multiple comparisons, all p-values equal 1.
This initial finding does not seem to align with our hypotheses, where we might expect more pronounced differences in the performance between layers.
To disentangle these findings, we conduct more thorough analyses between the layers below in Section~\ref{rq:layers}.

\begin{figure*}[htbp!]
    \centering
    \includegraphics[width=\textwidth]{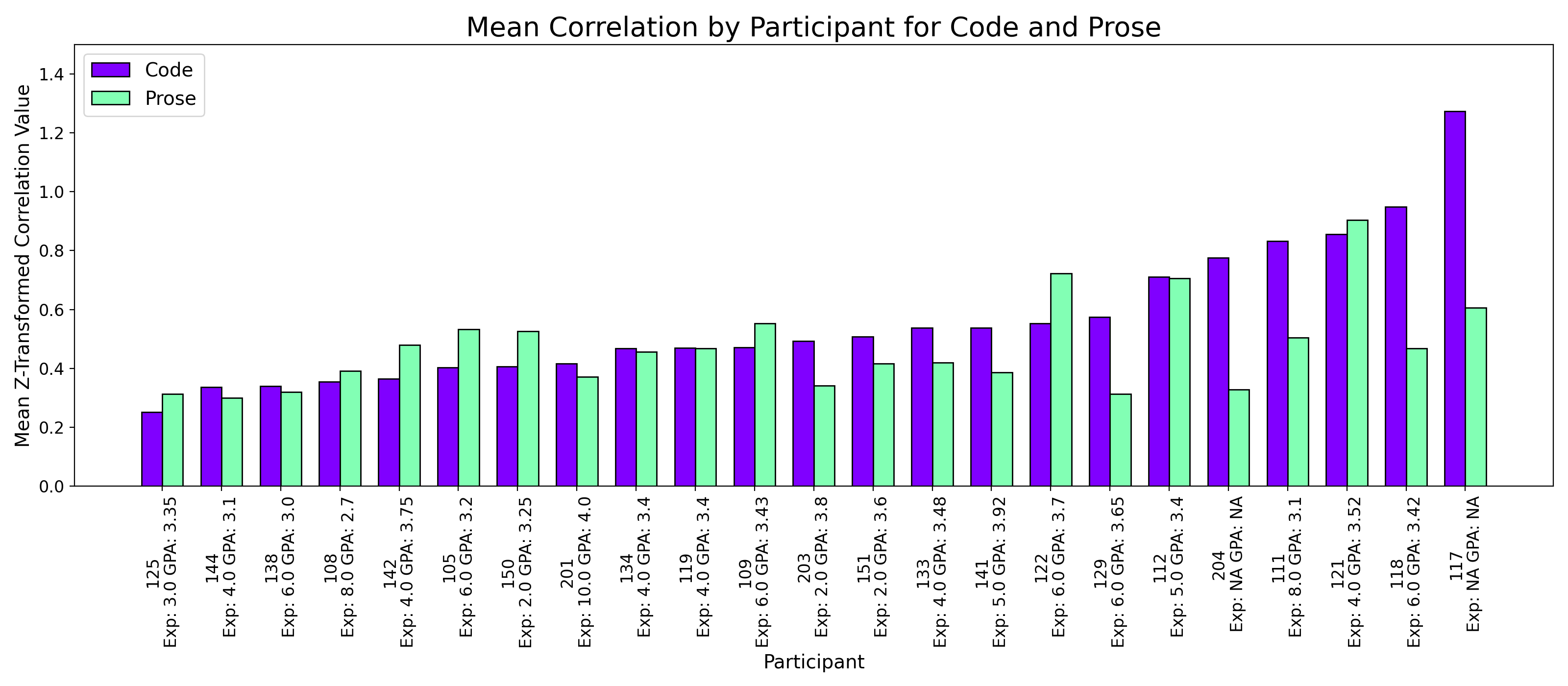}
    \caption{Modeling performance for each participant for the Code and Prose tasks, along with their years of coding experience (Exp.), and Grade Point Average (GPA). Modeling performance was calculated as the average of the top 10K z-transformed correlation coefficients for each participant and each of the layers, models, and parameter combination we considered.}
    \label{fig:participants_bar}
\end{figure*}

We next compared modeling performance between participants for Code and Prose to broadly understand individual differences.
For each participant, we computed the average z-transformed correlation coefficient among the top 10K voxels across the various layers, models, and parameter configurations, separately for Code and Prose. 
Results can be seen in Figure~\ref{fig:participants_bar}, where in the majority of participants, we observe higher performance in modeling brain activity while writing code compared to writing prose. 
This result approaches statistical significance, according to a paired t-test between participants' average correlation coefficients for Code and Prose ($t=1.987$, $p=0.060$, $d=0.424$, $\mu_{Code}=0.560$, $\sigma_{Code}=0.234$, $\mu_{Prose}=0.471$, $\sigma_{Prose}=0.149$).
Based on the figure, there does not appear to be a strong relationship between participants' correlation coefficients for Code and Prose, or between modeling performance and other factors like years of coding experience or GPA. 
Furthermore, we did not see large differences in modeling performance above between different LLM layers, but we do see a large degree of individual variability between participants.

\begin{center}
\fbox{\begin{minipage}{\textwidth}
Across a wide range of parameter combinations, the voxelwise BOLD signal for code writing can be predicted significantly better than that for prose writing ($p<0.001$, FDR-corrected). 
We find no significant group-level differences in how well different LLM layers predict participants' voxelwise BOLD signal during code and prose writing.
At the subject level, there is no significant difference in performance for predicting participants' voxelwise BOLD signal during Code or Prose.
\end{minipage}}\end{center}

\subsection{Task Performance and Model Alignment}\label{rq:performance}
Our initial line of questioning uncovered general trends comparing VEM performance between Code and Prose, but we next sought to gain a deeper understanding into behavioral and experimental factors that may influence modeling performance.
We also assessed how modeling performance related to participants' performance on the tasks, and whether we could better predict participants' brain activity if they performed well on the tasks.
In other words, have LLMs and high-performing programmers arrived at a shared cognitive representation for solving computational problems?
For simplicity, we focus all subsequent analyses on a single parameter configuration from above.
Specifically, we constrained our analyses to correlation coefficients from DeepSeek 6B that include 10 delayed copies of the feature vectors and no look ahead volumes. 
We used this parameter configuration for both Code and Prose because it performs the best for the former, and to keep all else consistent in our comparisons between Code and Prose. 
This configuration is best for the values of parameters we considered, but there may be another configuration that is globally optimum that we did not find.
For this reason, we prioritized controlling for parameter configurations between Code and Prose.
For completeness, we report key findings below using parameter configurations from other models in Table~\ref{tab:other_llms} in Section~\ref{rq:layers}. 

\textbf{Modeling Performance Correlates.} To investigate factors that may relate to modeling performance for Code and Prose, we conducted Pearson Correlation tests between participants' modeling performance, measured as the average of the top 10K z-transformed correlation coefficients for each participant, and other variables of interest.
Specifically, we considered participants' age, years of coding experience, and Grade Point Average (GPA) to understand whether participants' brain activity could be better predicted if they were more proficient with coding or had been coding for longer.
We also tested whether modeling performance might relate to participants' behaviors inside the scanner, such as the number of keystrokes they typed, the number of questions they got correct, or the amount that they edited their answers.
Here we operationalized editing behavior as the total number of backspaces divided by the total number of keystrokes, which we calculated separately for Code and Prose.
This may be a naive metric for editing behavior, especially since prior research has investigated editing while typing\citep{tian2026linking}, which we discuss below in Section~\ref{sec:discussion}.
Nonetheless, we hypothesized that participants' brain activity may not align well with LLM embeddings if they switch between different strategies while answering a question.

\begin{figure*}[htbp!]
    \centering
    \includegraphics[width=\textwidth]{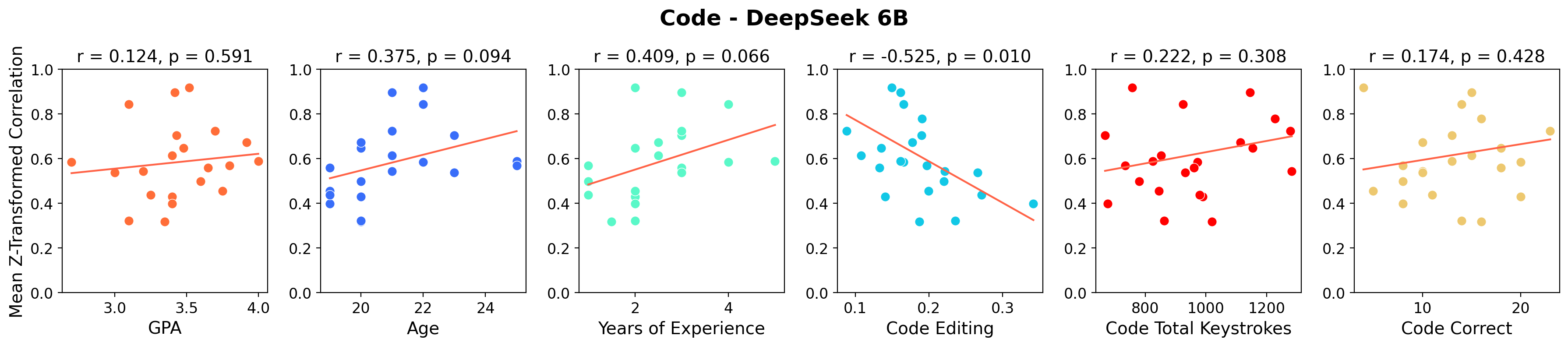}
    \caption{Correlation between variables of interest and participants' top 10K z-transformed correlation coefficients for Code for the top-performing model. Code Editing was operationalized by dividing the number of backspaces a participant typed by the total number of keystrokes they typed, and Code Correct is a measure of the number of coding questions the participants answered correctly.}
    \label{fig:code_corr}
\end{figure*}

\begin{figure*}[htbp!]
    \centering
    \includegraphics[width=\textwidth]{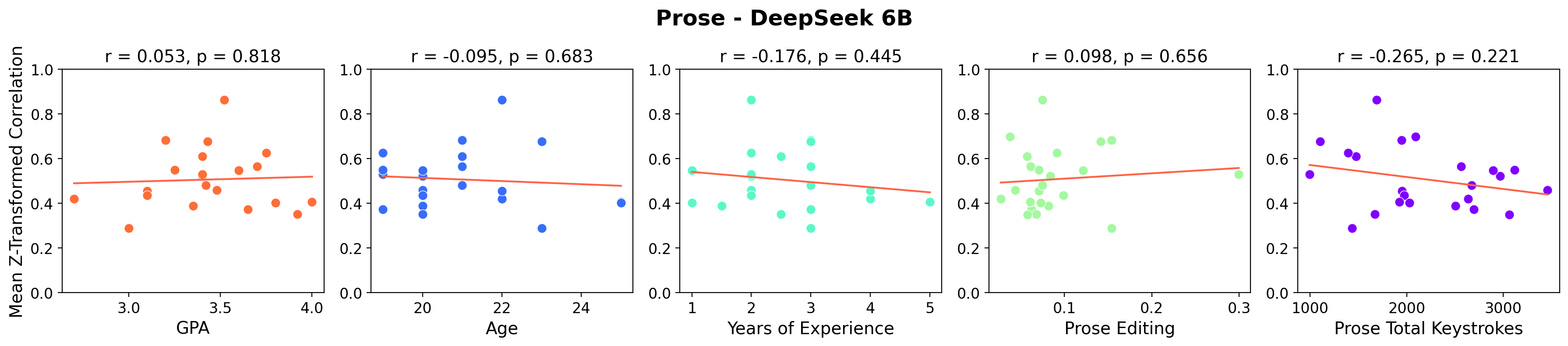}
    \caption{Correlation between participants' top 10K z-transformed correlation coefficients for Prose for the top-performing model and parameter combination and different variables of interest. The Prose task was subjective, so there were no measures of performance for this task.}
    \label{fig:prose_corr}
\end{figure*}

For Code, we considered participants' performance on the coding tasks, which is measured as the number of coding questions they got correct, as defined by the original study authors~\citep{krueger2020neurological}.
Since the Prose task was subjective, there was not a metric for performance that we considered for this condition.
Results for correlation tests between modeling performance and variables of interest for Code and Prose can be seen in Figures \ref{fig:code_corr} and~\ref{fig:prose_corr}, respectively.
We found a significant correlation only for modeling performance for Code with participants' code editing behaviors ($r=-0.525$, $p=0.010$).
In other words, as participants edited the code more, the performance decreased for modeling their brain activity based on the code they wrote.
We found that the correlation between years of experience and modeling performance approached significance ($r=0.409$, $p=0.066$).
Our statistical power may not be strong enough for this test, but this trend may suggest that more time spent coding may lead to stronger alignment between LLM embeddings and a human's neural representation for generating code.
The weaker relationship between modeling performance and other variables of interest suggest that modeling performance using participants' keystrokes may not strongly relate to more stable facets of cognition like years of experience, but may relate more to finer-scale behaviors inside of the scanner.
Notably, there are some participants who performed well on the task (i.e., Participant 108), but whose brain activity could not be predicted well from the LLM embeddings, which has interesting implications that we discuss further in Section~\ref{sec:discussion}.

\textbf{Brain Regions.} 
From our results so far, we have seen evidence that there is individual variability in how well participants' brain activity can be modeled using LLM embeddings.
To understand where in the brain these differences may manifest, we split our participants into groups, then compared them based on the top-modeled regions for participants within that group.
Specifically, we tested for Code and Prose whether there were differences in top-modeled regions based on (1) participants' performance on the tasks, and (2) how well participants' voxelwise BOLD signal was modeled.
Separately for Code and Prose, we split participants into two groups based on model alignment, as depicted in Figure~\ref{fig:participants_bar}, and the number of questions they got correct in the coding condition.
We did not have a metric for participants' performance on the Prose condition since this task was more subjective, so we split participants by code performance for this comparison as well.

For the coding task, after splitting participants into two roughly equally-sized groups, we assigned 12 participants to the High Performing and Well Modeled groups, and 11 participants to the Low Performing and Poorly Modeled groups.
There were eight participants that were common between the High Performing and Well Modeled groups, meaning that the group compositions were 33\% different.
We also found above that modeling performance did not significantly correlate with task performance.
For Prose, we again assigned 12 participants to the High Performing and Well Modeled groups, and 11 participants to the Low Performing and Poorly modeled groups, with only six participants that were common between Well Modeled and High Performing.
Between the Well Modeled groups for Code and Prose, there were only six participants that were common between both.

We identified the brain regions where most top 10K voxels were located, which was based on the correlation strength between the predicted and recorded BOLD signal. 
For these 10K voxels, we identified their corresponding parcel number in the Schaefer Atlas, and tallied the parcel numbers that occurred most often. 
This yielded a ranked list for each participant of Schaefer parcels that contained the highest number of voxels among the top 10K.
The ranked order of these parcels could potentially be influenced by the size of the parcels, where larger parcels that contain more voxels could be overrepresented. 
We considered raw counts here to understand where these top 10K voxels are located, but include results in Supplementary Material about \textit{top-modeled regions} where we normalize counts by region size (Section~\ref{supp}).
For raw counts, we hypothesized that regions that have previously been found to be relevant for code and prose writing, such as Broca's Area, the Number Form Area, and parietal regions, to be locations for many top 10K voxels for participants~\citep{krueger2020neurological, karas2021connecting, peitek2021program}.
For participants who were modeled poorly and performed poorly on the tasks, we hypothesized that their top 10K voxels would be more randomly distributed throughout the brain.

\begin{figure*}[htbp!]
    \centering
    \includegraphics[width=\textwidth]{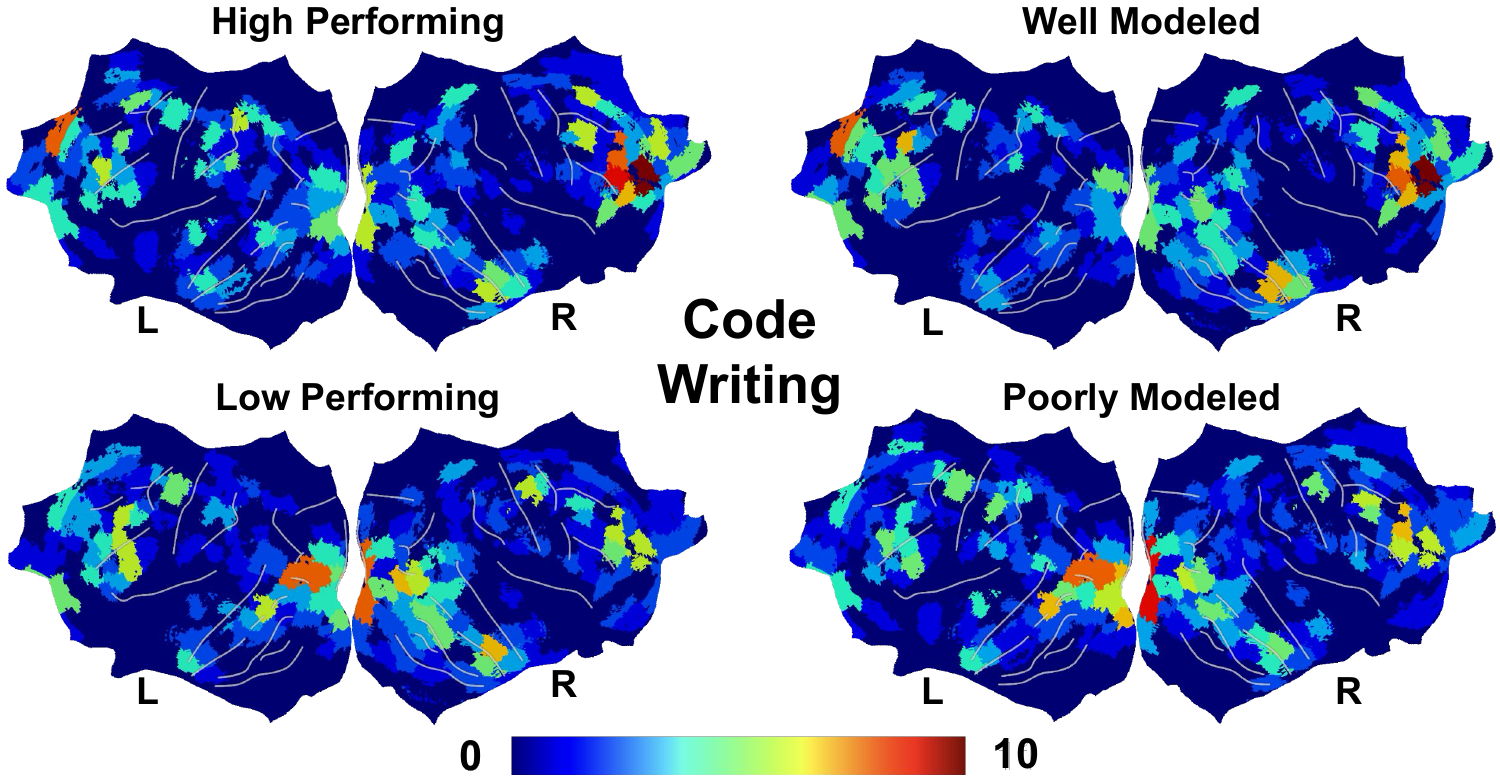}
    \caption{Tallies in each Schaefer parcel for the number of participants for whom it was among the top 25 locations for best-modeled parcels for Code. We ranked parcels for each participant by counting the number of voxels among the top 10K that were within the parcel. We split participants into High Performing ($n=12$) and Low Performing ($n=11$) groups based on their performance during the coding task, and Well Modeled ($n=12$) and Poorly Modeled ($n=11$) based on the average of their top 10K z-transformed correlation coefficients for the Code task (Figure~\ref{fig:participants_bar}). There were eight participants that were common between the High Performing and Well Modeled groups.}
    \label{fig:code_brains}
\end{figure*}

\begin{figure*}[htbp!]
    \centering
    \includegraphics[width=\textwidth]{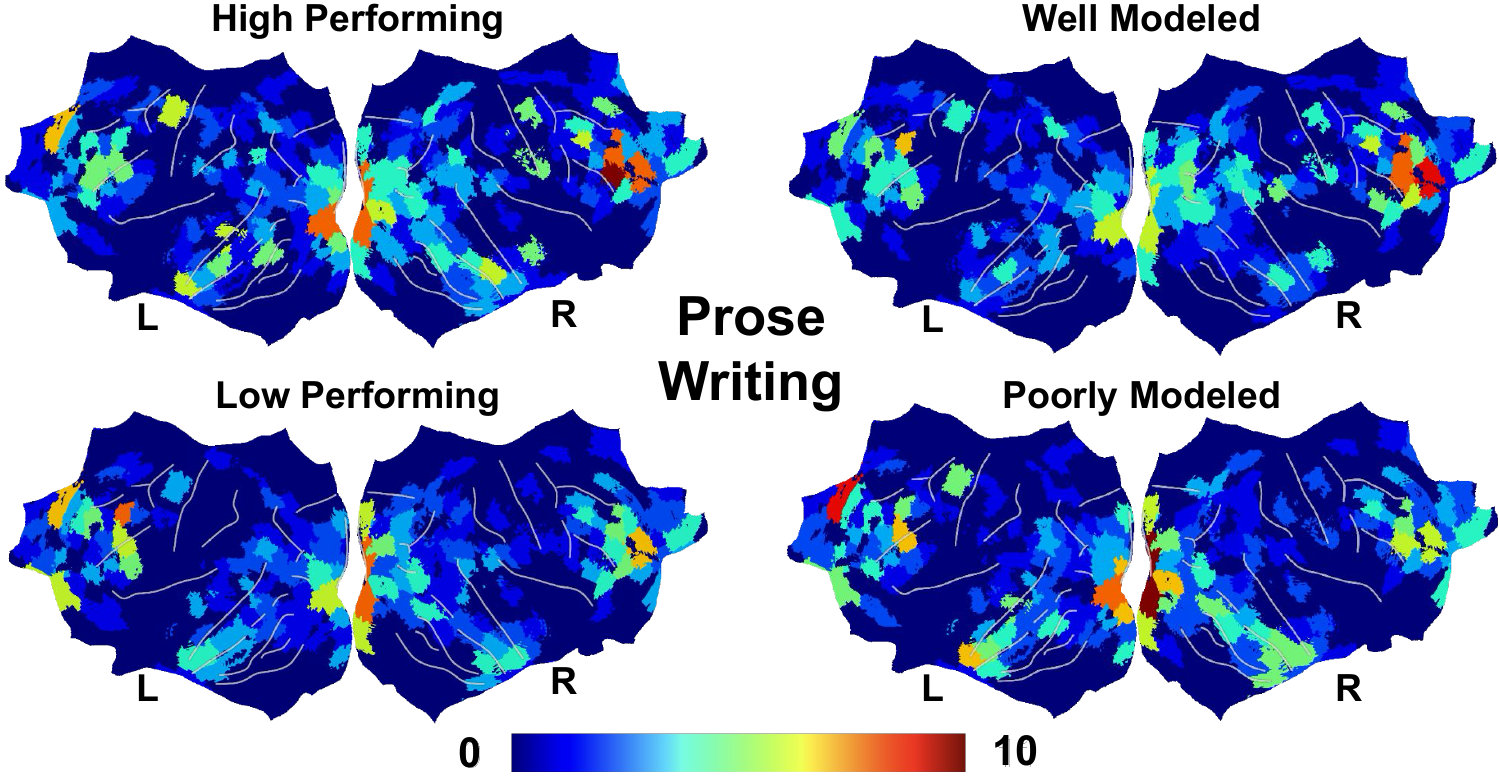}
    \caption{Tallies in each Schaefer parcel for the number of participants for whom it was among the top 25 locations for best-modeled voxels for Prose. We split participants into High Performing ($n=12$) and Low Performing ($n=11$) groups based on their performance during the coding task, and Well Modeled ($n=12$) and Poorly Modeled ($n=11$) based on the average of their top 10K z-transformed correlation coefficients for the Prose task (Figure~\ref{fig:participants_bar}). There were six participants that were common between the High Performing and Well Modeled groups.}
    \label{fig:prose_brains}
\end{figure*}

To visualize group-level patterns for top voxel locations, we counted the number of participants for whom a given Schaefer parcel was among the top 25 locations for top 10K voxels\footnote{We considered the top 25 schaefer parcels because we also use the Harvard-Oxford atlas in Section~\ref{rq:layers} to complement the Schaefer Atlas for high-level interpretability, and there are on average 4.54 Schaefer parcels in each Harvard-Oxford region. We thus considered the top 25 regions for consistency.}.  
We tallied these counts separately for Code (Figure~\ref{fig:code_brains}) and Prose (Figure~\ref{fig:prose_brains}). 
From Figure~\ref{fig:code_brains}, we observe striking consistency in top regions between participants who performed well on the coding task, and those were modeled well using LLM embeddings of their keystrokes. 
Notably, the regions that are consistently among the top regions for participants in both the High Performing group and Well Modeled group are prominently located in the right frontal pole.
Researchers have found that this region is involved in delayed intentions (i.e., prospective memory)~\citep{costa2013right}, multitasking~\citep{roca2011role}, and information integration~\citep{chau2025complex}, which may suggest that high-level information processing strategies may be shared between LLMs and the human brain.
For the Low Performing and Poorly Modeled groups, we see that regions in the lateral occipital cortex are consistently among participants' top voxel locations.
This suggests that LLM embeddings may be limited to modeling low-level visual processes in participants who are low performing and poorly modeled. 
We see these patterns for Prose as well, where the right frontal pole is consistently among the top voxel locations for participants in the High Performing and Well Modeled groups.
This suggests that these regions may involve general process that are shared between code and prose writing, which is supported by previous research~\citep{chau2025complex, costa2013right}.

\begin{center}
\fbox{\begin{minipage}{\textwidth}
Focusing on the best performing model and combination of parameters (DeepSeek 6B with 10 delayed copies of the feature vectors and no ``look-ahead'' volumes), there is a significant negative correlation between modeling performance for Code and participants' amount of code editing ($r=-0.525$, $p=0.010$).
The right frontal pole was the most frequent location for top-modeled voxels in participants whose brain activity was modeled well. 
This was true for both Code and Prose, and also for participants who performed well on the coding task, even though the group compositions were 33\% to 50\% different.
The positive correlation between modeling performance and years of coding experience approached significance ($r=0.409$, $p=0.066$).
\end{minipage}}\end{center}

\subsection{Layer Investigation}\label{rq:layers}
Given the evidence from previous research suggesting that certain LLM layers process different linguistic features~\citep{lopez2025linguistic}, we analyzed whether embeddings from different LLM layers were better able to model activity in different brain regions.
We focused on the DeepSeek 6B model with the highest performing parameter configuration (Section~\ref{rq:parameters}), from which we extracted embeddings from eight layers evenly spaced throughout the model. 
We again considered the top 10K voxels whose activity were best modeled using these embeddings, and used the Harvard-Oxford Atlas to enable high level interpretation~\citep{makris2006decreased, jenkinson2012fsl}.
To this end, we identified corresponding regions for the top 10K voxels in the Harvard-Oxford Atlas based on their coordinates.
Similar to above, here we consider raw counts in Harvard-Oxford Atlas regions, but in Supplementary Material we include results that normalize counts by region size to investigate top-modeled regions (Section\ref{supp}).
As described in Section~\ref{sec:methods}, we computed VEMs for voxels in the Schaefer Atlas, but the two atlases do not align precisely.
Thus, we omitted voxels among the top 10K that did not have corresponding regions in the Harvard-Oxford Atlas ($\mu=715.2$ voxels omitted).
We then counted the number of voxels among the remaining top 10K that were located within each region, and analyzed differences in these top locations between the various LLM layers.

\begin{figure*}[htbp!]
    \centering
    \begin{subfigure}[t]{0.48\textwidth}
        \centering
        \includegraphics[width=\textwidth]{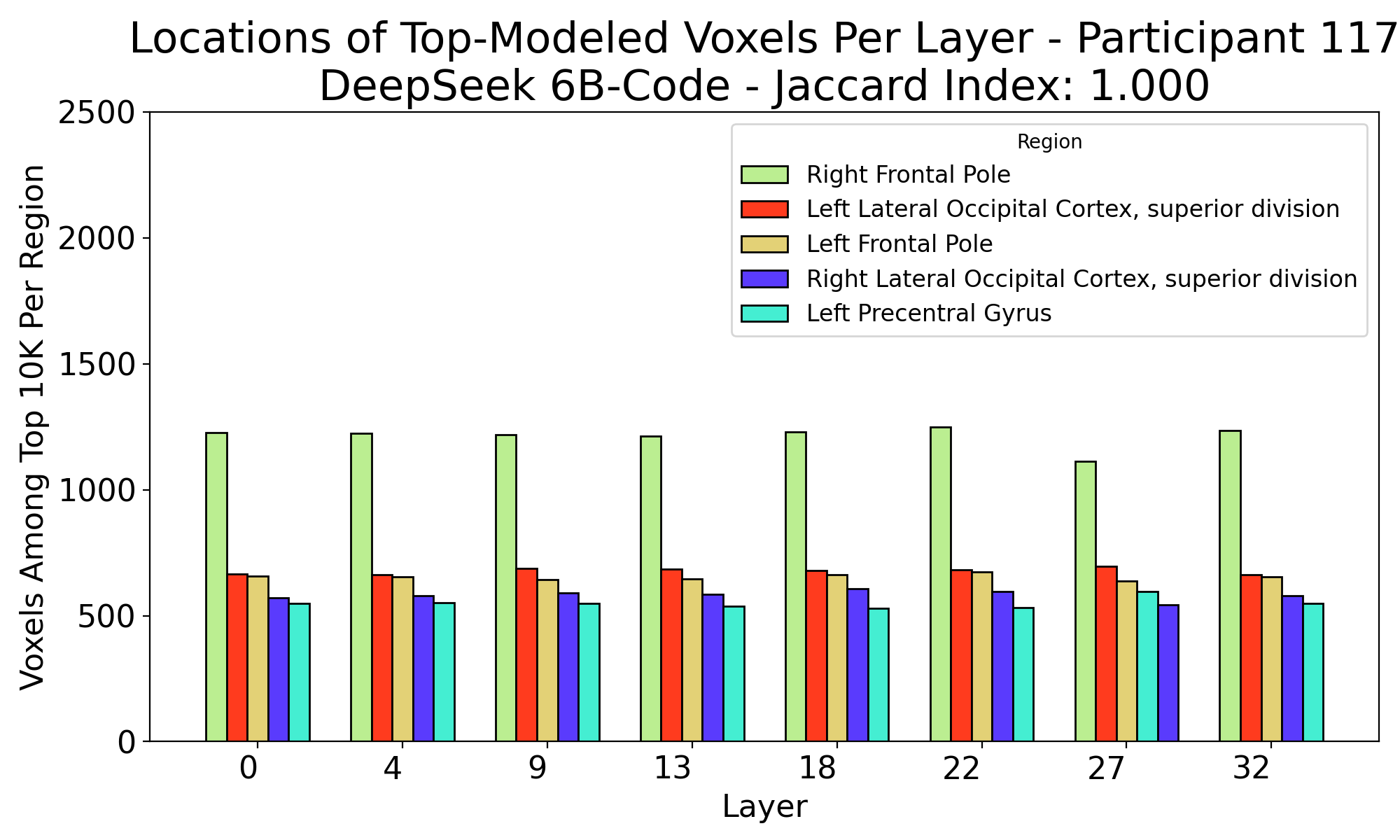}
        \caption{}
    \end{subfigure}
    \hfill
    \begin{subfigure}[t]{0.48\textwidth}
        \centering
        \includegraphics[width=\textwidth]{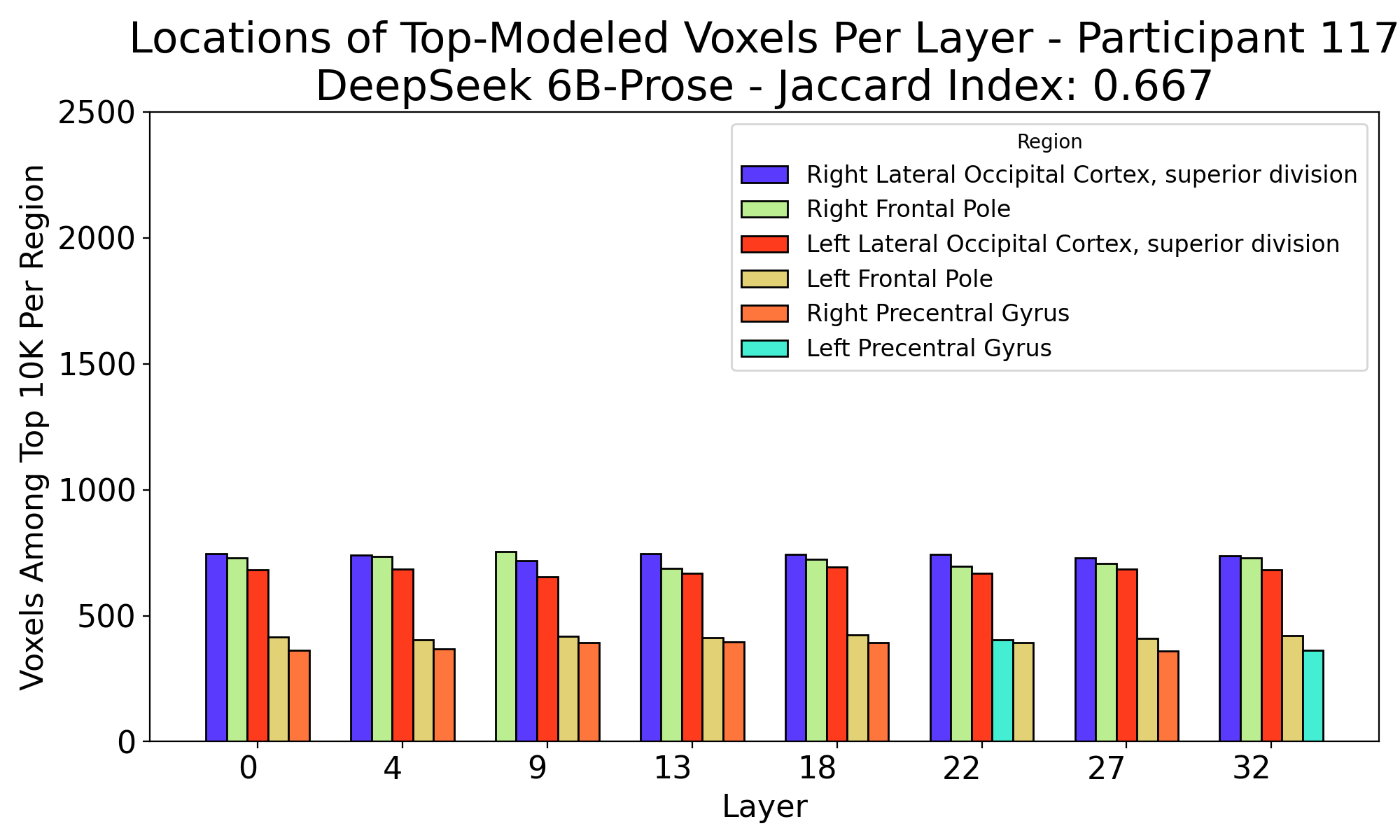}
        \caption{}
    \end{subfigure}
    \caption{Locations of the top 10K modeled voxels in the Harvard-Oxford Atlas, filtered to the top five regions for each layer for the best performing model and parameter configurations. Depicted here is Participant 117 for (a) Code and (b) Prose, whose z-transformed correlation coefficients were highest for Code among all the participants. The Jaccard Index is calculated by dividing the intersection of different sets by their union, and here is the number of the regions that are shared between all the layers divided by the total number of regions present across the layers. This serves as a measure of the similarity in the top regions across the layers.}
    \label{fig:top_bar}
\end{figure*}

\begin{figure*}[htbp!]
    \centering
    \begin{subfigure}[t]{0.48\textwidth}
        \centering
        \includegraphics[width=\textwidth]{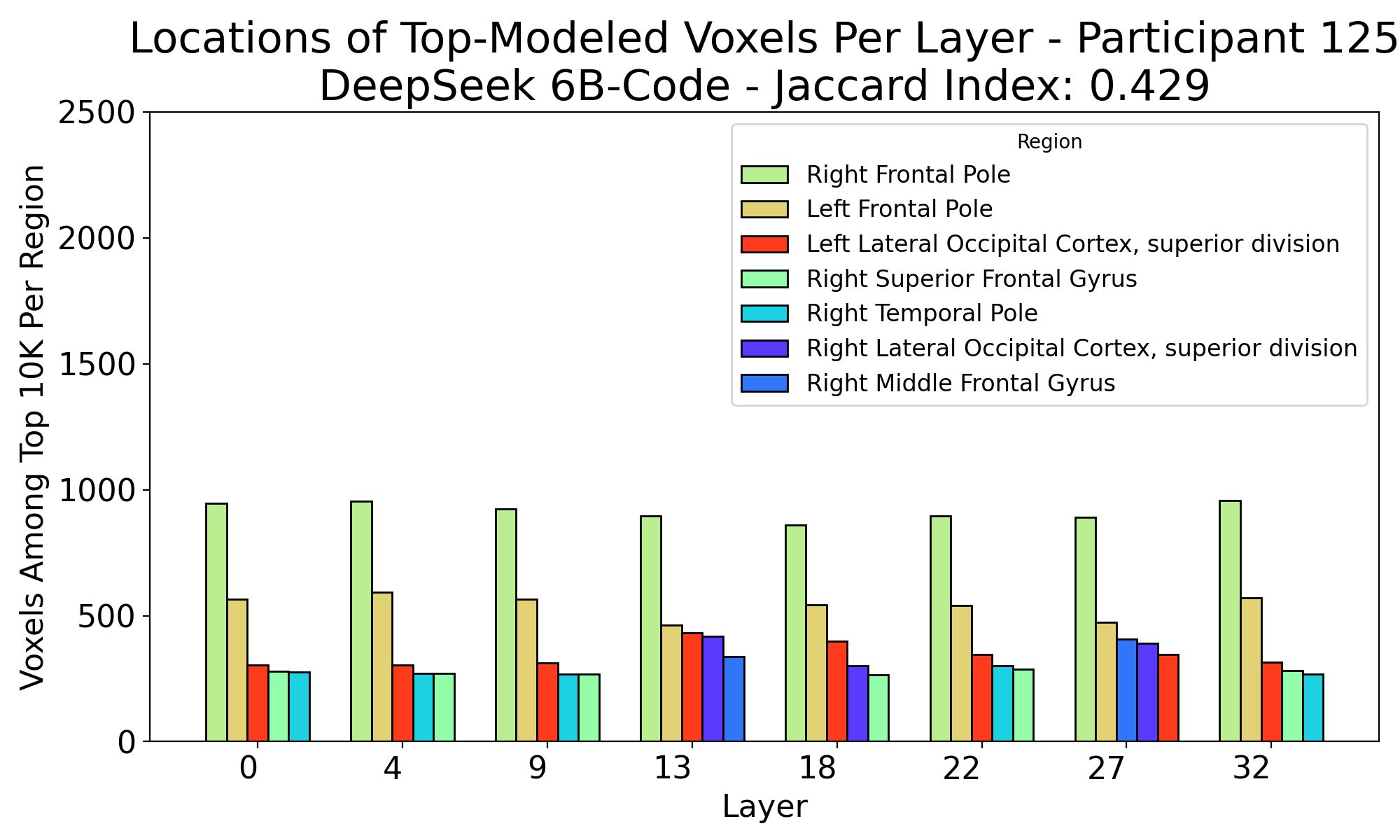}
        \caption{}
    \end{subfigure}
    \begin{subfigure}[t]{0.48\textwidth}
        \centering
        \includegraphics[width=\textwidth]{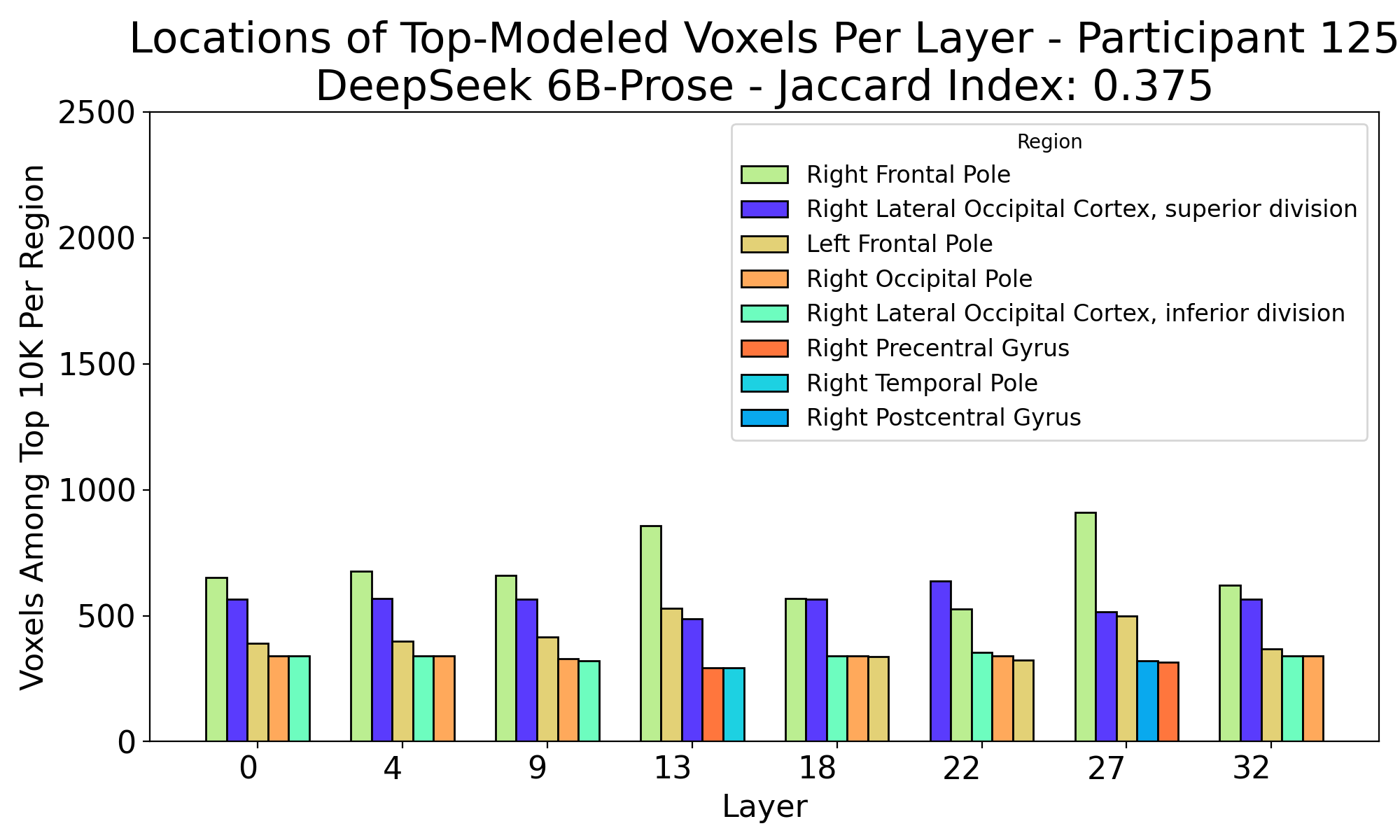}
        \caption{}
    \end{subfigure}
    \caption{Locations of the top 10K modeled voxels in the Harvard-Oxford Atlas, filtered to the top five regions for each layer for the best performing model and parameter configurations. Depicted here is Participant 125 for (a) Code and (b) Prose, whose z-transformed correlation coefficients were lowest for Code among all the participants.}
    \label{fig:bottom_bar}
\end{figure*}

Bar graphs of voxel counts among the top 10K within Harvard-Oxford regions for each layer can be seen in Figure~\ref{fig:top_bar}, which depicts the participant whose voxelwise BOLD signal was predicted best, on average, across layers and parameter configurations.
Figure~\ref{fig:bottom_bar} depicts the participant whose voxelwise BOLD signal was predicted worst.
From these figures, we observe a striking degree of consistency in the top voxel locations per layer. 
Considering Participant 117 for Code, for instance, the same five regions are among the top five voxel locations for each of the layers we considered throughout the model. 
We observe this consistency to a lesser degree for Prose, and for both tasks for Participant 125.
We test below whether, across participants, the amount of consistency across layers has a significant relationship with modeling performance.

\begin{figure*}[htbp]
    \centering
    \includegraphics[width=\textwidth]{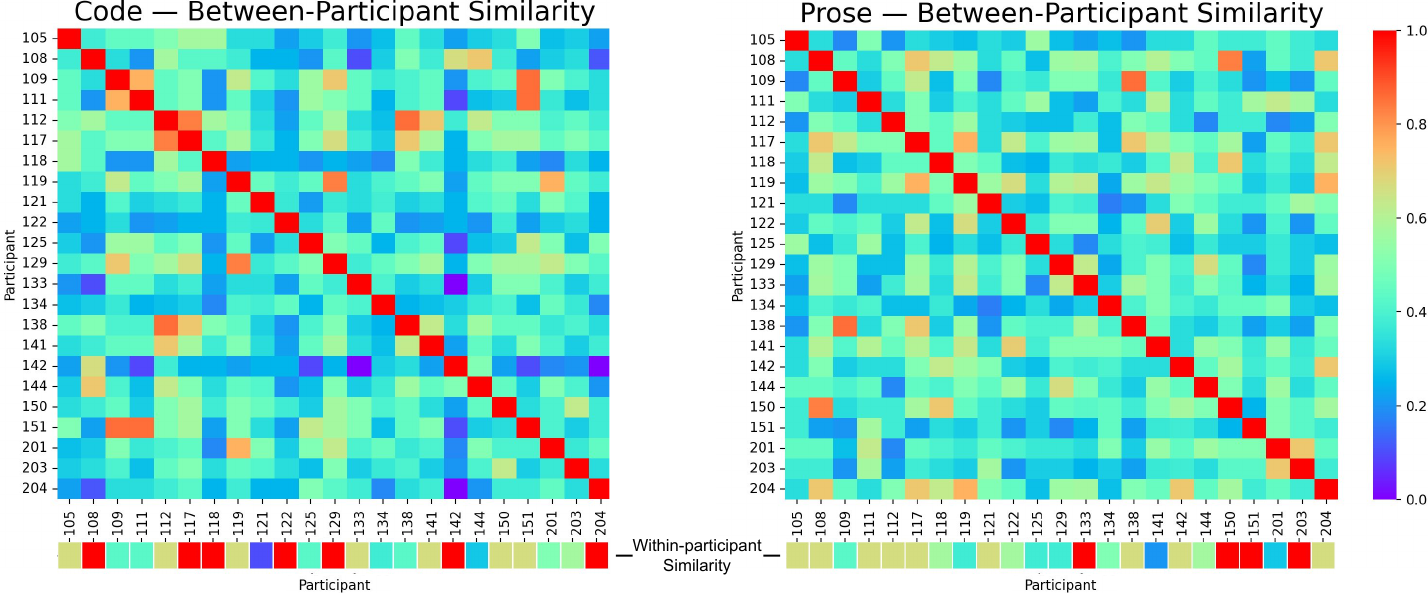}
    \caption{Pairwise similarities in the top regions between all participants. These values were calculated by computing the Jaccard Index between two given participants' top-modeled regions. These sets of regions for each participant were computed by taking the union of their top-modeled regions across the different LLM layers. Also depicted is the within-participant similarity, which is the Jaccard Index across the different model layers for each participant.}
    \label{fig:similarity}
\end{figure*}

To measure the consistency across the layers for a given participant and task, we computed the Jaccard Index, which is a widely used measure from set theory that measures the degree of overlap between sets, which has applications in Natural Language Processing and image processing, among others~\citep{costa2013right}.
Specifically, the Jaccard Index is a ratio that divides the intersection of the sets by the union of the sets.
Thus, for our purposes, if the top five brain regions across the layers demonstrate a large degree of overlap, the number of regions that are shared between the layers will be closer to the number of total regions across the layers. 
We see that the average Jaccard Index is higher for Code ($\mu=0.661$, $\sigma=0.263$) than it is for Prose ($\mu=0.620$, $\sigma=0.221$), which is depicted in the Within-participant Similarity at the bottom of Figure~\ref{fig:similarity}.
This difference was not statistically significant, according to a paired t-test ($t=0.586$, $0.564$).
From Figure~\ref{fig:bottom_bar}, we observe that participant 125 exhibits more variability in the regions that were modeled best across the model layers compared to Participant 117, so we tested whether this was a broader trend across participants. 
Specifically, we tested whether modeling performance correlates with the variability of brain regions modeled by each LLM layer, but did not find a significant correlation between the Jaccard Index and modeling performance for Code ($r=0.257$, $p=0.236$) or Prose ($r=0.114$, $p=0.604$). 

While there is a high degree of consistency between layers, we also observe that the set of top regions across the layers is quite distinct for a given participant and task (i.e., the regions modeled for Code and Prose for Participant 117 in Figure~\ref{fig:top_bar}). 
This suggests that each participant may have a unique ``fingerprint'' in how their brain activity aligns with LLM embeddings.
To test the degree of similarity between participants' top voxel locations, we calculated the pairwise Jaccard Index between each participant.
Specifically, we computed the set of all the top five regions across the eight layers for a given participant and task (i.e., Participant 117 on Prose: Right Lateral Occipital Cortex (superior), Right Frontal Pole, Left Lateral Occipital Cortex (superior), Left Frontal Pole, Right Precentral Gyrus, Left Precentral Gyrus, Figure~\ref{fig:top_bar}).
Using these sets of regions for each participant and task, we then computed the pairwise Jaccard Index between every participant, which is depicted in Figure~\ref{fig:similarity}.

The average similarity between participants' top regions for Code is 0.388 ($\sigma = 0.149$) and for Prose is 0.414 ($\sigma = 0.137$), as measured by the Jaccard Index. 
This finding indicates that the top voxel locations are only about 39\% similar between participants for Code, and only 41\% similar for Prose.
Using a pairwise t-test, we found that this difference between Code and Prose was statistically significant ($t=2.109$, $p<0.05$). 
This suggests that the top voxel locations across participants are more consistent for prose writing than for code writing. 
This may be due to the fact that participants had written prose for longer, so there may be higher specialization and stability in brain regions that are involved~\citep{kuhl2010brain, liu2025rapid}.
More broadly, the relatively low similarity between participants may indicate that the representation encoded within the LLM embeddings is uniquely distributed across the brain for a given participant and task.

\begin{table}[htbp]
    \small
    \centering
    \caption{Within- and between-participant similarities based on Jaccard Index for other top-performing parameter configurations for Code and Prose, as well as the top-modeled voxel locations for the other LLMs. Results align with our findings for DeepSeek 6B, with the exception of CodeGemma 7B.}
    \label{tab:other_llms}
    \resizebox{\textwidth}{!}{
    \begin{tabular}{l|l|r|r|r|r|p{8cm}}
        \hline
        \makecell[l]{\textbf{Task}} & \makecell[l]{\textbf{Model}} & \makecell[r]{\textbf{Number of}\\ \textbf{Delayed Copies}} & \makecell[r]{\textbf{Look-Ahead}\\\textbf{Volumes}} & \makecell[r]{\textbf{Within-Participant}\\ \textbf{Similarity}} & \makecell[r]{\textbf{Between-Participant}\\ \textbf{Similarity}} & \makecell[l]{\textbf{Top-Modeled Voxel Locations}} \\
        \hline
        \multirow{5}{*}{Code} & \multirow{5}{*}{DeepSeek 6B} & \multirow{5}{*}{10} & \multirow{5}{*}{0} & \multirow{5}{*}{$0.661\pm0.263$} & \multirow{5}{*}{$0.388\pm0.149$} & \begin{enumerate}[noitemsep, topsep=7pt, parsep=0pt, partopsep=0pt, leftmargin=*, before=\vspace{-\baselineskip}]  \item Right Frontal Pole
          \item Left Frontal Pole
          \item Right Lateral Occipital Cortex, superior division
          \item Left Lateral Occipital Cortex, superior division
          \item Right Precentral Gyrus
        \end{enumerate}\\\hline
        \multirow{5}{*}{Code} & \multirow{5}{*}{CodeGemma 7B} & \multirow{5}{*}{10} & \multirow{5}{*}{1} & \multirow{5}{*}{$0.381\pm0.246$} & \multirow{5}{*}{$0.389\pm0.140$} & \begin{enumerate}[noitemsep, topsep=7pt, parsep=0pt, partopsep=0pt, leftmargin=*, before=\vspace{-\baselineskip}]  \item Right Frontal Pole
          \item Left Frontal Pole
          \item Right Lateral Occipital Cortex, superior division
          \item Left Lateral Occipital Cortex, superior division
          \item Left Precentral Gyrus
        \end{enumerate}\\\hline
        \multirow{5}{*}{Code} & \multirow{5}{*}{StarCoder 3B} & \multirow{5}{*}{10} & \multirow{5}{*}{0} & \multirow{5}{*}{$0.552\pm0.235$} & \multirow{5}{*}{$0.422\pm0.148$} & \begin{enumerate}[noitemsep, topsep=7pt, parsep=0pt, partopsep=0pt, leftmargin=*, before=\vspace{-\baselineskip}]  \item Right Frontal Pole
          \item Left Frontal Pole
          \item Left Lateral Occipital Cortex, superior division
          \item Right Lateral Occipital Cortex, superior division
          \item Right Precentral Gyrus
        \end{enumerate}\\\hline
        \multirow{5}{*}{Prose} & \multirow{5}{*}{StarCoder 7B} & \multirow{5}{*}{16} & \multirow{5}{*}{3} & \multirow{5}{*}{$0.560\pm0.203$} & \multirow{5}{*}{$0.408\pm0.156$} & \begin{enumerate}[noitemsep, topsep=7pt, parsep=0pt, partopsep=0pt, leftmargin=*, before=\vspace{-\baselineskip}]  \item Left Frontal Pole
          \item Right Frontal Pole
          \item Right Lateral Occipital Cortex, superior division
          \item Left Lateral Occipital Cortex, superior division
          \item Right Precentral Gyrus
        \end{enumerate}\\\hline
        \multirow{5}{*}{Prose} & \multirow{5}{*}{CodeGemma 7B} & \multirow{5}{*}{16} & \multirow{5}{*}{3} & \multirow{5}{*}{$0.460\pm0.233$} & \multirow{5}{*}{$0.400\pm0.136$} & \begin{enumerate}[noitemsep, topsep=7pt, parsep=0pt, partopsep=0pt, leftmargin=*, before=\vspace{-\baselineskip}]  \item Right Frontal Pole
          \item Left Frontal Pole
          \item Right Lateral Occipital Cortex, superior division
          \item Left Lateral Occipital Cortex, superior division
          \item Right Precentral Gyrus
        \end{enumerate}\\\hline
        \multirow{5}{*}{Prose} & \multirow{5}{*}{DeepSeek 6B} & \multirow{5}{*}{16} & \multirow{5}{*}{3} & \multirow{5}{*}{$0.646\pm0.199$} & \multirow{5}{*}{$0.375\pm0.152$} & \begin{enumerate}[noitemsep, topsep=7pt, parsep=0pt, partopsep=0pt, leftmargin=*, before=\vspace{-\baselineskip}]  \item Right Frontal Pole
          \item Left Frontal Pole
          \item Right Lateral Occipital Cortex, superior division
          \item Left Lateral Occipital Cortex, superior division
          \item Left Occipital Pole
        \end{enumerate}\\\hline
    \end{tabular}
    }
\end{table}

For completeness, we also computed key metrics for other LLMs that we considered in our study.
Specifically, in Table~\ref{tab:other_llms} we report the between- and within-participant similarities, as well as the top voxel locations, for the best-performing parameter configurations in the LLMs we considered in this study.
We see a large degree of consistency between the LLMs in demonstrating that within-participant similarity ($\mu=0.543$, $\sigma=0.108$) is higher than between-participant similarity ($\mu=0.397$, $\sigma=0.017$), which is statistically significant according to a pairwise t-test
($t=3.188$, $p<0.05$, $d=1.301$).
These results demonstrate that our findings based on DeepSeek 6B generalize to the other state of the art LLMs as well, with CodeGemma 7B as an exception.

\begin{center}
\fbox{\begin{minipage}{\textwidth}
For a given participant, the brain regions where most top-modeled voxels were located 66\% the same across LLM layers for code writing and 62\% the same for prose writing.
The brain regions best modeled by the LLM embeddings were only 39\% and 41\% the same between two given participants, on average, for code and prose writing, respectively.
These trends that within-participant similarity is higher than between-participant similarity generalize to most state of the art LLMs we considered.
\end{minipage}}\end{center}

\section{Discussion}\label{sec:discussion}
In this study, we modeled human programmers' voxelwise BOLD signal using LLM embeddings of their keystrokes during code and prose writing tasks. 
We conducted thorough investigations to understand how well this modeling approach performs, the factors that may relate to modeling performance, regions in the brain that align well with LLM embeddings, and individual variability between participants.
Here we synthesize and interpret our findings, and discuss their implications as well as future directions. 

\subsection{Interpretations and Implications}
Our results demonstrate the potential and shortcomings for using LLM embeddings of participants' code and prose writing behavior to model their brain activity. 
Specifically, our key findings suggest that (1) modeling BOLD signal based on keystrokes is more feasible for code writing compared to prose writing, (2) activity in the right frontal pole consistently aligns well with LLM embeddings for participants who both perform well on the task and whose brain activity is well-modeled, and (3) the top-modeled regions for a given participant and task are consistent across LLM layers, yet distinctive between participants.

\textbf{Code versus Prose.} Broadly speaking, modeling brain activity of code writing may be easier than modeling prose due to code's regularity and structure, compared to the flexibility and expressivity of prose. 
This is supported by an influential paper in software engineering, where Hindle et al. demonstrated that the regularity and predictability of software is much higher than that of natural language~\citep{hindle2016naturalness}.
In this context, our findings support the hypothesis that writing processes for code are represented in a continuous numerical space in the human brain and LLMs, though this space may be more sparse for code than that of natural language.

\textbf{Right Frontal Pole.} We found consistent evidence that BOLD signal in the right frontal pole could be modeled well using LLM embeddings of participants' keystrokes.
This was especially pronounced for participants in the High Performing and Well Modeled groups, which suggests that cognitive representations in this region may converge towards those of LLMs for participants based on their coding strategies.
The right frontal pole has been causally implicated with Transcranial Magnetic Stimulation (TMS) in delayed intentions (i.e., prospective memory) during visual-spatial tasks, where participants were asked to recreate a pattern of dots on a computer screen~\citep{costa2013right}.
While coding does indeed involve visual-spatial processing~\citep{huang2019distilling}, we see that participants in both the Code and Prose conditions exhibit high modeling performance in this region, which suggests that alignment in the right frontal pole with LLM embeddings may not be specific to coding.

Research has also found evidence suggesting that the right frontal pole is involved in multitasking~\citep{roca2011role}, the selection and maintenance of high-order goals~\citep{roca2011role}, information integration~\citep{chau2025complex}, as well as synthesizing and reducing multiple sources of information into simpler features~\citep{chau2025complex}.
Together, those studies provide a framework for interpreting our findings, where both code and prose writing may involve prospective memory, integration of multiple sources of information, and maintenance of high-order goals.
Specific to this study where we model BOLD signal using LLM embeddings, these findings suggest that LLMs may perform similar processes internally, and that the activity in the right frontal pole can be grounded in the keystrokes participants typed.
In other words, this region has been implicated in high-order tasks, and we find that its activity can be modeled effectively based on participants' writing behaviors.
This provides an exciting research direction to untangle the activity in this area using features from participants keystrokes. 

We originally hypothesized that top-modeled voxels would be located primarily in inferior frontal and parietal regions implicated in prior research~\citep{peitek2021program, karas2021connecting}, which was not strongly supported by our findings.
The top-modeled regions for some participants were located in our hypothesized regions, but this broad divergence between our hypotheses and results may suggest that the ability to model activity in a brain region using LLM embeddings may not speak to its importance for a given task.
In other words, LLM embeddings may be an effective tool for modeling brain activity in a given region only if its cognitive representation is inherently close to that of an LLM.
For example, the BOLD signal in Broca's Area may reflect neural processes that are dissimilar from the LLM representations of participants' writing behaviors, so even though Broca's Area may be important for an activity like writing, its activity may not align well with an LLM.
This presents unique opportunities to study where and how neural representations are different from LLM representations, and may also suggest limitations for using LLMs to model brain activity.

\textbf{Layer Consistency.} Our analyses uncovered surprising evidence that the brain regions modeled best by various layers throughout the LLMs we considered were strikingly consistent. 
This finding is at odds with previous research that has convincingly reported how different LLM layers reflect different stages of natural language processing, such as processing lexical or semantic features in text~\citep{vaidya2022self, tenney2019bert}. 
Research that has investigated the information processed at different Transformer layers has largely used BERT models that are outperformed by current LLMs~\citep{lopez2025linguistic}, so current models may have advanced such that representations are distributed across the layers.
This suggests that our intuitions about the information that is represented in each layer may need updating.
In the same way that we have moved beyond the concept that each brain region has a particular function, it may be an oversimplification that each layer performs a single function.
Notably, our findings originate from a writing task whereas previous research has involved tasks like listening or reading~\citep{huth2016natural, vaidya2022self, toneva2019interpreting}, so more careful study is certainly required.

That being said, coordination across layers may be critical for the LLM to output a cohesive response, which may explain the prominence of the right frontal pole throughout our results.
This region is involved in the maintenance of high-order goals and information integration~\citep{roca2011role, chau2025complex}, so the alignment we found between LLM embeddings and voxels in the right frontal pole may suggest that LLMs are similarly maintaining high-order goals and synthesizing information across layers.
Prior research has theorized and found evidence that writing in the human brain is not a serial, step-by-step process, but more parallelized and collaborative between different prefrontal, parietal, and occipital regions~\citep{becker2006review, chen2019delineating}.
Our results support these findings and suggest that LLMs may use a similar, parallel process. 
In contrast to BERT models, modern LLMs may have advanced past being a pipeline that sequentially processes linguistic information. 


\textbf{Participant Variability.} Lastly among our interpretations, the fine-tuning process of LLMs has been found to drastically limit the variability of their responses.
This has been termed \textit{mode collapse}, where models will output virtually the same text to a prompt each time~\citep{o2024attributing}.
This may be desirable for commercial products, but may be detrimental for using LLM embeddings to model BOLD signal for a diverse sample of human participants.
Thus, the homogeneity of LLM responses may limit the usage of LLM embeddings to model the brain activity of participants who use different coding strategies.
This may explain why there were participants performed well on the task, but whose brain activity was poorly modeled. 
Studying these participants in particular may uncover distinct patterns of cognition that provide an alternative perspective on solving computational problems.

\subsection{Future Directions}
Our findings modeling BOLD signal from LLM embeddings based on participants' keystrokes can influence future research, education, and software engineering practices, and can also inform AI model training.
Returning to our original study aim to improve our understanding of both cognition and AI tools with respect to coding,
our findings inform where brain activity and LLM embeddings may share common representations, and where they may diverge. 
Our findings strongly suggest that compared to prose writing, code writing is more aligned between the human brain and LLMs.
Broadly, this closer alignment for code writing compared to prose writing can be leveraged to design future AI tools that \textit{predict} code, but \textit{assist} with prose.
For example, neurally-aligned AI tools for programming education could help students arrive at an optimal solution, and step in when they stray from expected patterns. 
This can be extended to tools used in software engineering as well, where models could help identify when developers deviate from project-specific guidelines.
By contrast, AI tools for prose writing may be naturally suited to collaborate with students as they organize their thoughts, develop their vocabulary, and structure their arguments.

Considering opportunities for future research, we found that modeling performance was not significantly correlated with factors like performance on the task or GPA, but was instead significantly correlated with factors like code editing behavior. 
We considered metrics like code editing behavior and total keystrokes to characterize participants' typing behaviors inside of the scanner, and there may be other behaviors inside the scanner that can be used to more robustly contextualize the BOLD signal.
One possibility could be to add context using participants' gaze behavior recorded using eye-tracking, which provides an additional measure of cognition~\citep{sharafi2015systematic}.
Overall, our study has demonstrated the potential to use LLMs to model brain activity during code and prose writing tasks, but our approach can be used to inform future research that similarly studies generative processes such as speaking or social interactions.
Analogous to using participants' keystrokes, in these contexts researchers could potentially use transcripts of what participants said to model brain activity.

\subsection{Limitations} The limitations of this study also present opportunities for future research.
For instance, we relied on participants' keystrokes, but there were windows of time when participants did not type.
Without information for these timepoints, the VEM's estimation of BOLD signal were static.
We attempted to mitigate this limitation by removing timepoints during the rest periods, so our analyses focused on relevant time windows where possible.
Future research could consider other sources of information, such as eye tracking, to augment the model of the BOLD signal during those timepoints.
Moreover, Huth et al. analyzed the features of the ridge regression model to understand each voxel's tuning towards different semantic features.
The ridge regression models in our study could similarly be used to understand how voxels are tuned to features in the code.
The editing behaviors we considered in this study were admittedly rudimentary, dividing the number of backspaces by the total number of keystrokes, but researchers have studied editing behaviors during writing~\citep{tian2026linking}, which can be leveraged here to develop a stronger understanding of editing behavior that may relate to brain activity.

Furthermore, the code questions were only given in the C++ programming language, and the prose questions were only administered in English. 
Future research might consider additional programming and natural languages, but we nonetheless found significant differences between Code and Prose with implications for education and AI-tool design.
The tasks were also administered in the MRI scanner, which could have influenced participants' performance and comfortability, especially considering participants were required to type without seeing the keyboard. 
This had a notable influence on some participants' performance who could not touch type, and may contribute to our results about participants' whose BOLD signal could not be predicted well from their keystrokes.
These findings are still informative for understanding the factors that may influence modeling performance, but future studies should consider factors like touch typing ability during recruitment.

\section{Conclusions}
In this study, we used LLM embeddings of participants' keystrokes during code and prose writing tasks to predict their brain activity. 
We found that these embeddings could be used to predict the activity better for code writing compared to prose writing, which was especially pronounced in the right frontal pole (a region implicated in high level goal planning, prospective memory, and information integration). 
This has implications for AI-tools that more reliably predict code completion as opposed to supporting natural language tasks.
We also found that for a given participant, the brain regions best modeled by LLM layers spaced throughout the model were 66\% the same for code writing and 62\% the same for prose writing.
\textit{Between} two given participants, however, the brain regions best modeled by the LLM embeddings were only 39\% and 41\% the same for code and prose writing, respectively.
This consistency within participants suggests that information may be represented across layers within modern LLMs.
The variability across participants suggests that each individual may have a unique ``fingerprint'' in how their brain activity can be modeled using LLM embeddings.
Overall, our study finds that LLM embeddings of participants' keystrokes are a viable tool for studying cognition associated with code and prose writing.

\section*{Data and Code Availability}

Neuroimaging data used in the study is openly available upon request, and analysis code can be found at \url{https://github.com/largehappygroup/fmri_model}.
Any inquiries are welcome and can be directed to the corresponding author at z.karas@vanderbilt.edu.

\section*{Author Contributions}

Zachary Karas: study conceptualization, data preprocessing and analysis, results interpretation, manuscript preparation;
Catie Chang - domain expertise, study conceptualization, project guidance, results interpretation;
Kevin Leach - study conceptualization, project guidance;
Yu Huang - study conceptualization, project guidance, results interpretation.
AI tools were minimally used in preparation of the manuscript for editing and rephrasing text, and initial rough drafting.



\section*{Declaration of Competing Interests}

The authors report no competing interests. 

\section*{Acknowledgements}
We express our sincere gratitude to Tyler Santander for his guidance and advice during data analysis, and to Jerry Tang for directing us to a voxelwise encoding model tutorial.

\section*{Supplementary Material}\label{supp}

\textbf{Additional Results: Top-modeled Brain Regions.}
In Sections~\ref{rq:performance} and~\ref{rq:layers}, we created ranked lists of brain regions where most top 10K voxels were located.
These results could be biased towards larger brain regions that contain more voxels, so we sought to identify brain regions whose activity may be more specialized for code or prose writing.
As such, we performed the same analyses but first normalized the brain regions by the number of voxels therein.
Specifically, for a given region, we divided the number of voxels among the top 10K by the total number of voxels in that region.
We made this change for mapping top 10K voxels to both Schaefer parcels and Harvard-Oxford regions.
Here we similarly constrained our analyses to the best performing parameter combination of DeepSeek 6B (10 delayed copies, 0 look-ahead volumes), and in Table\ref{tab:other_llms_proportions} present key results from other LLMs in our study.
By normalizing voxel counts by region size, we are identifying the regions with the highest \textit{proportion} of top 10K voxels to investigate the regions that are modeled \textit{best} using LLM embeddings.

\begin{figure*}[htbp!]
    \centering
    \includegraphics[width=\textwidth]{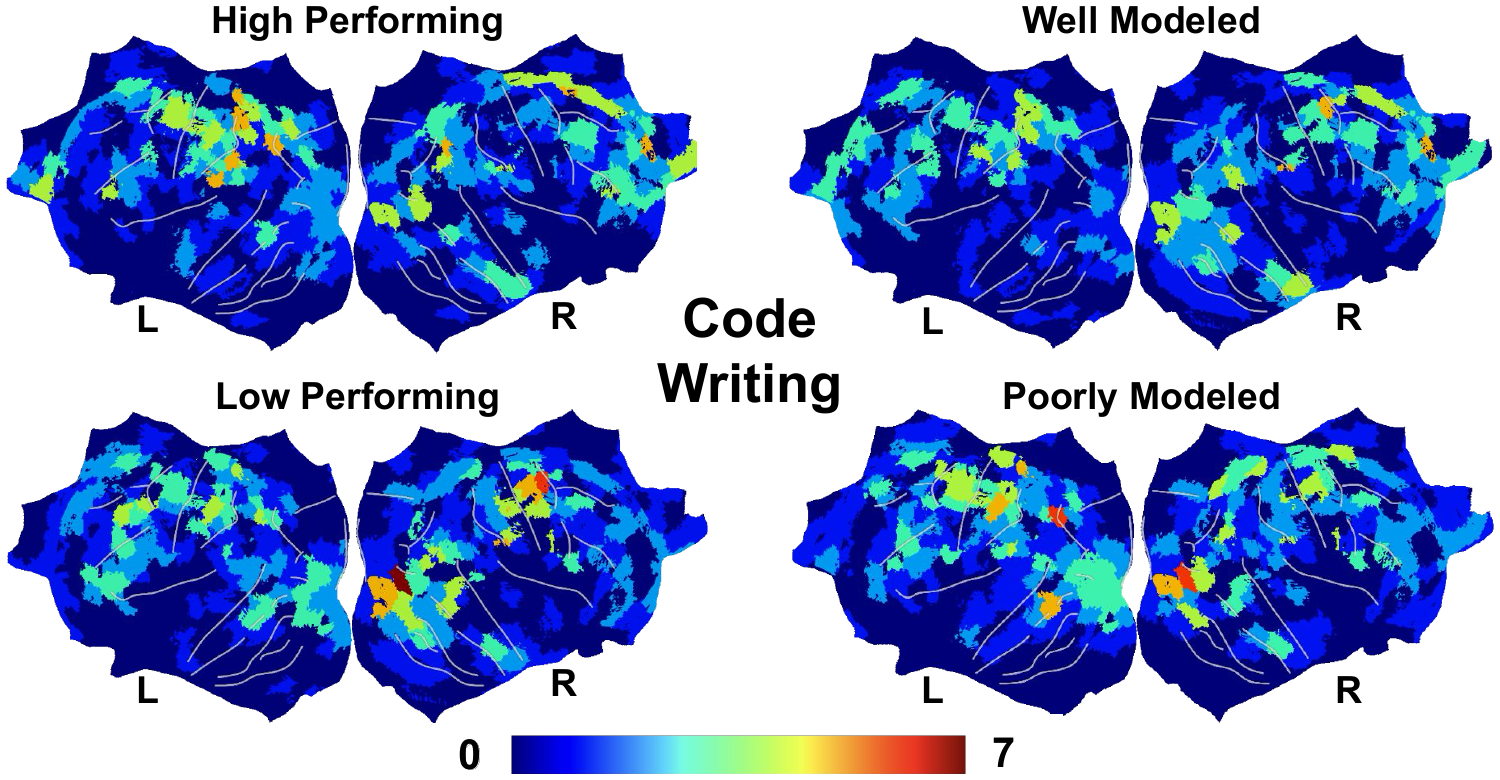}
    \caption{Tallies in each Schaefer parcel for the number of participants for whom it was among the top 25 best-modeled parcels for Code. We ranked parcels for each participant by counting the number of voxels among the top 10K that were within the parcel, then dividing by the total number of voxels in a given parcel. We split participants into High Performing ($n=12$) and Low Performing ($n=11$) groups based on their performance during the coding task, and Well Modeled ($n=12$) and Poorly Modeled ($n=11$) based on the average of their top 10K z-transformed correlation coefficients for the Code task (Figure~\ref{fig:participants_bar}). There were eight participants that were common between the High Performing and Well Modeled groups.}
    \label{fig:code_brains_proportions}
\end{figure*}

\begin{figure*}[htbp!]
    \centering
    \includegraphics[width=\textwidth]{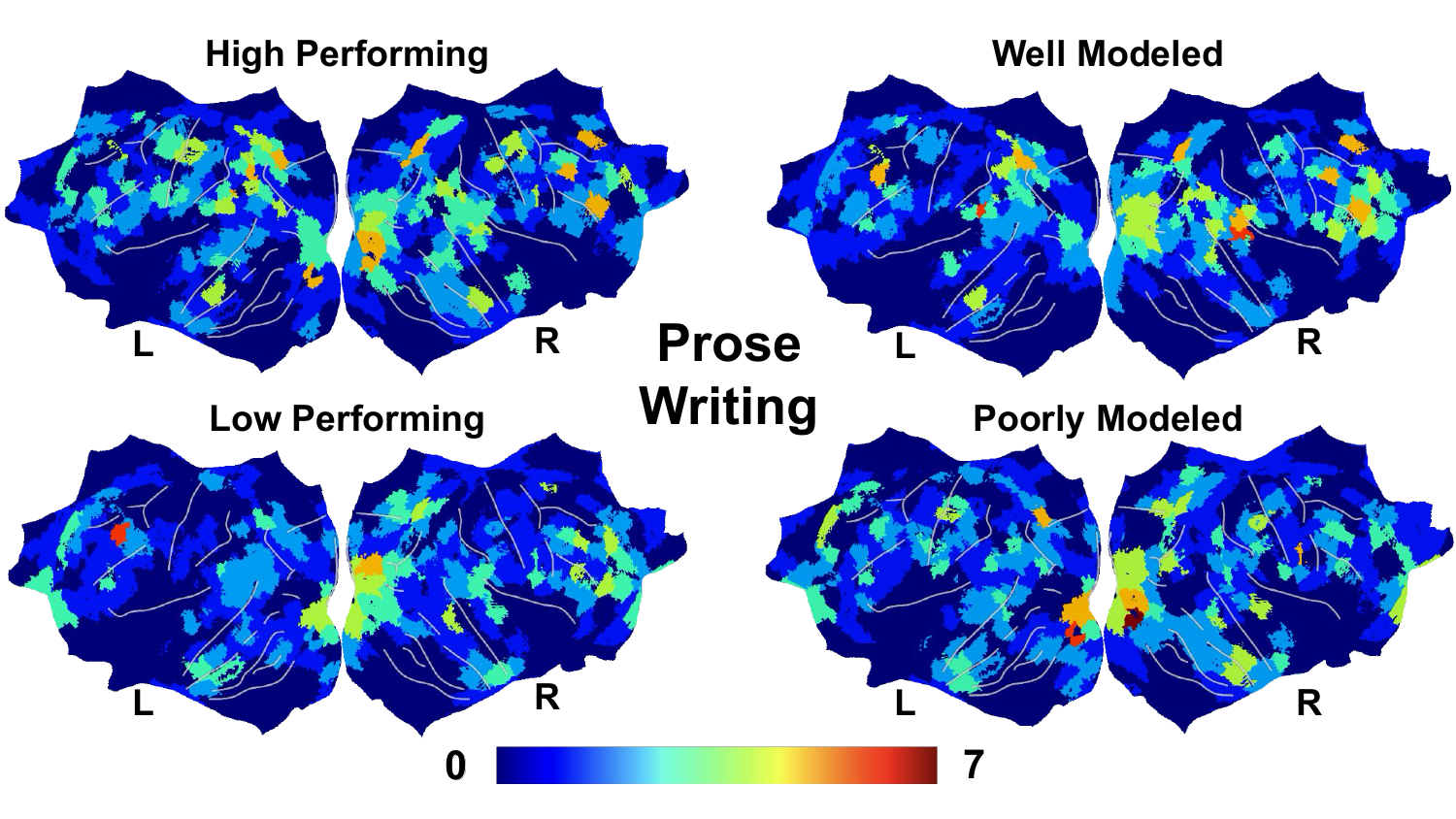}
    \caption{Tallies in each Schaefer parcel for the number of participants for whom it was among the top 25 best-modeled parcels for Prose. We split participants into High Performing ($n=12$) and Low Performing ($n=11$) groups based on their performance during the coding task, and Well Modeled ($n=12$) and Poorly Modeled ($n=11$) based on the average of their top 10K z-transformed correlation coefficients for the Prose task (Figure~\ref{fig:participants_bar}). There were six participants that were common between the High Performing and Well Modeled groups.}
    \label{fig:prose_brains_proportions}
\end{figure*}

We split our participants into groups based on how well their brain activity was modeled, and how well they performed on the coding task.
Our group compositions were consistent with those from above in Section~\ref{rq:performance}.
From Figure~\ref{fig:code_brains_proportions}, we observe that the top parcels are more distributed among participants in each group compared to raw counts in Figure~\ref{fig:code_brains}.
This is especially pronounced in left parietal regions and bilaterally in the motor cortex. 
We no longer see consistency between the High Performing and Well Modeled groups for code or prose writing, where top regions here are distributed across left parietal and right medial frontal areas.
Ranking regions by how well they are modeled may accentuate the varability between participants in how their brain activity aligns with LLM embeddings.
We see a similar trend for prose writing in Figure~\ref{fig:prose_brains_proportions}, where the top regions are more distributed across participants in each group compared to results when we use raw counts.
Broadly, we observe that top regions for participants in the High Performing and Well Modeled groups are clustered in parietal, occipital, and frontal regions.
For participants in the Low Performing and Poorly Modeled groups, however, top brain regions seem to primarily be clustered in occipital regions.
This suggests that for participants in these groups, LLM embeddings are more reliably able to predict brain activity in low-level sensory regions as opposed to higher-order regions involved in planning or spatial-temporal processing.

\begin{figure*}[htbp!]
    \centering
    \begin{subfigure}[t]{0.48\textwidth}
        \centering
        \includegraphics[width=\textwidth]{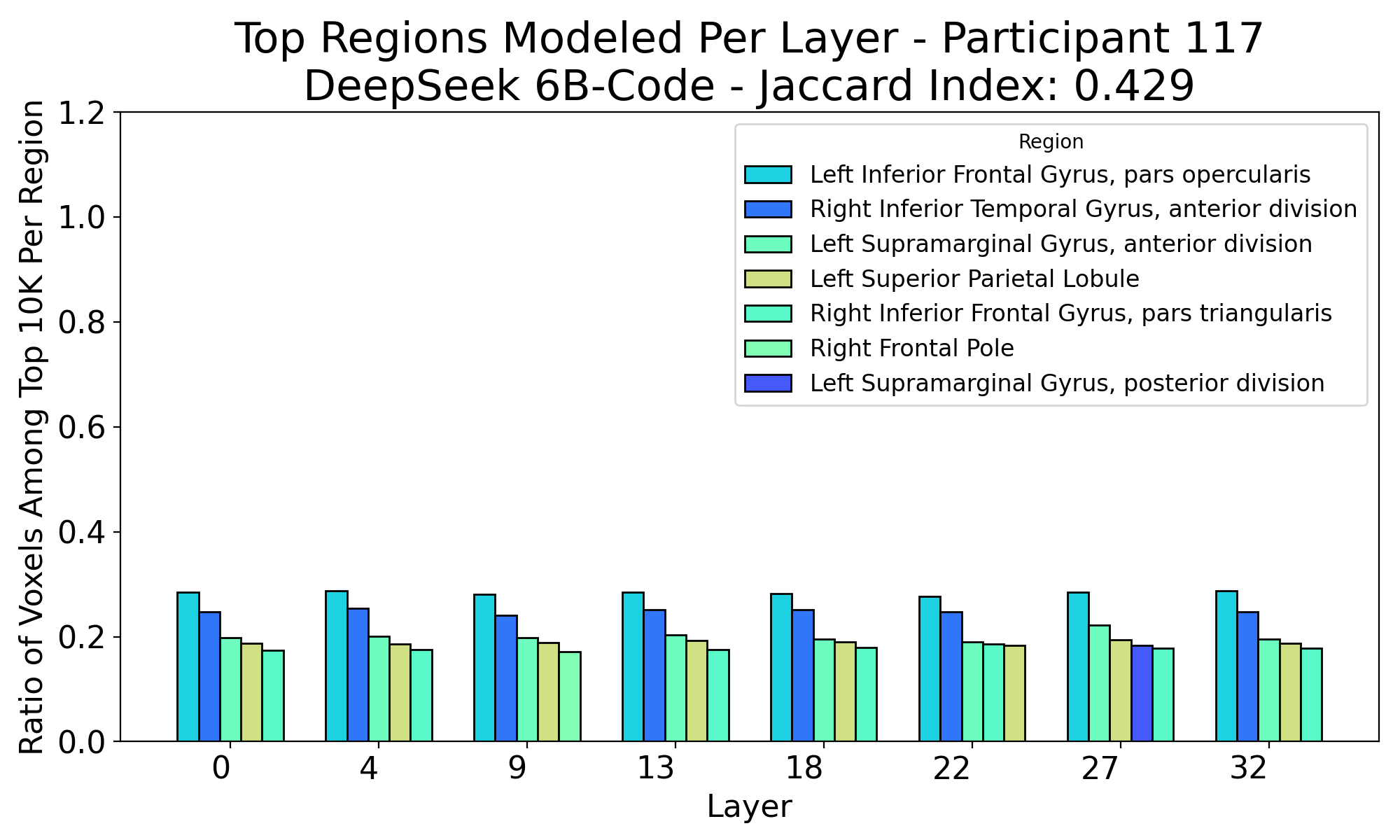}
        \caption{}
    \end{subfigure}
    \hfill
    \begin{subfigure}[t]{0.48\textwidth}
        \centering
        \includegraphics[width=\textwidth]{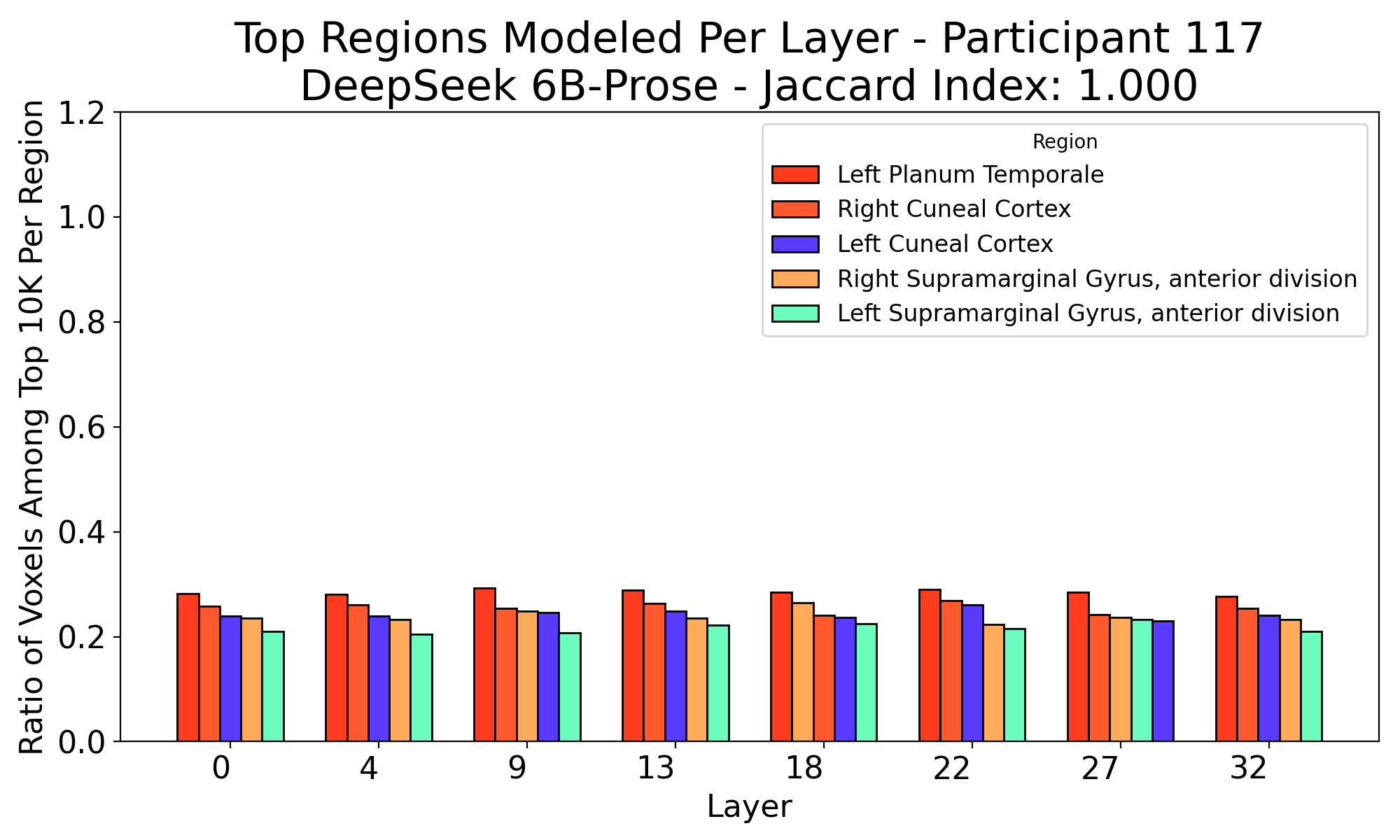}
        \caption{}
    \end{subfigure}
    \caption{Top-modeled regions in the Harvard-Oxford Atlas, filtered to the top five regions for each layer for the best performing model and parameter configurations. Depicted here is Participant 117 for (a) Code and (b) Prose, whose z-transformed correlation coefficients were highest for Code among all the participants.}
    \label{fig:top_bar_proportions}
\end{figure*}

\begin{figure*}[htbp!]
    \centering
    \begin{subfigure}[t]{0.48\textwidth}
        \centering
        \includegraphics[width=\textwidth]{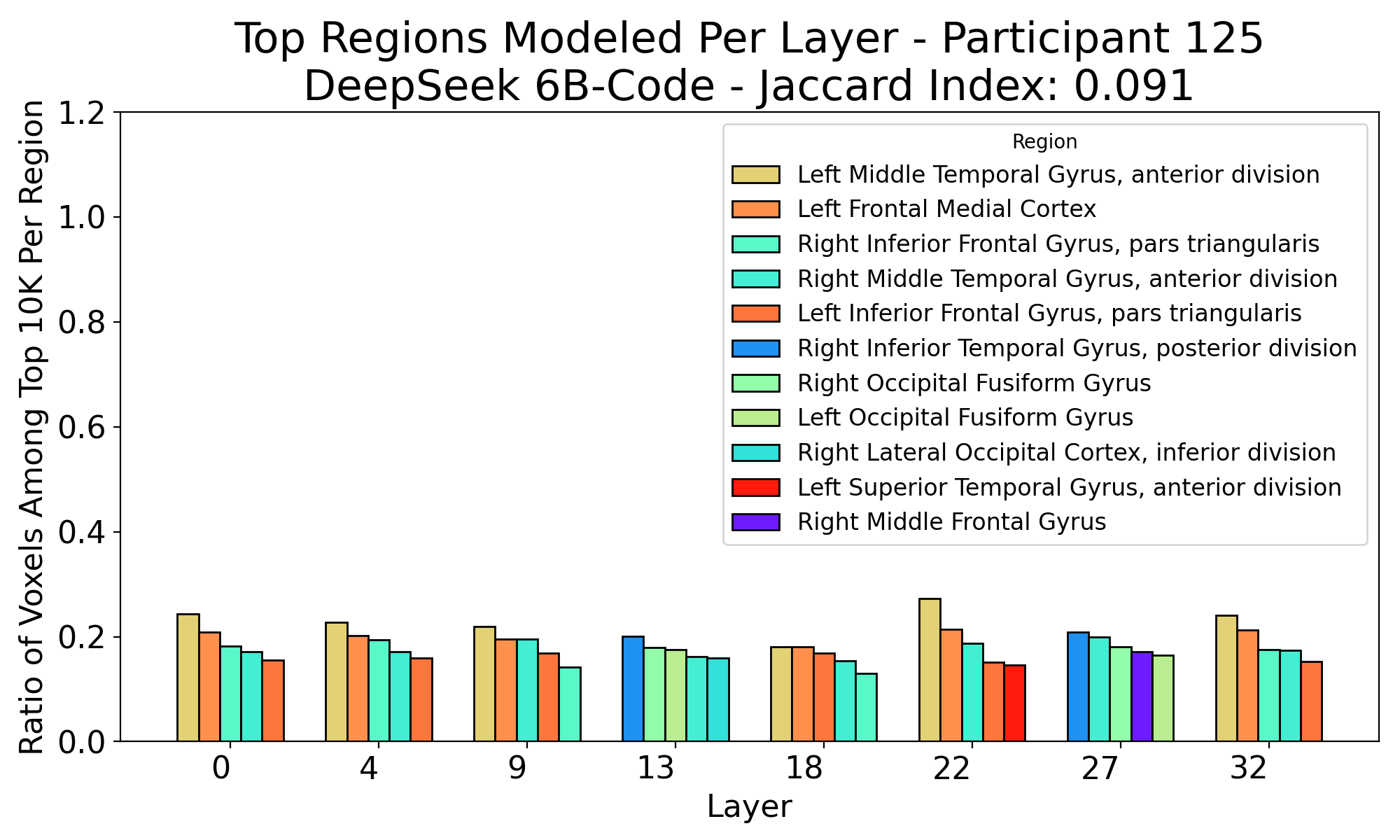}
        \caption{}
    \end{subfigure}
    \begin{subfigure}[t]{0.48\textwidth}
        \centering
        \includegraphics[width=\textwidth]{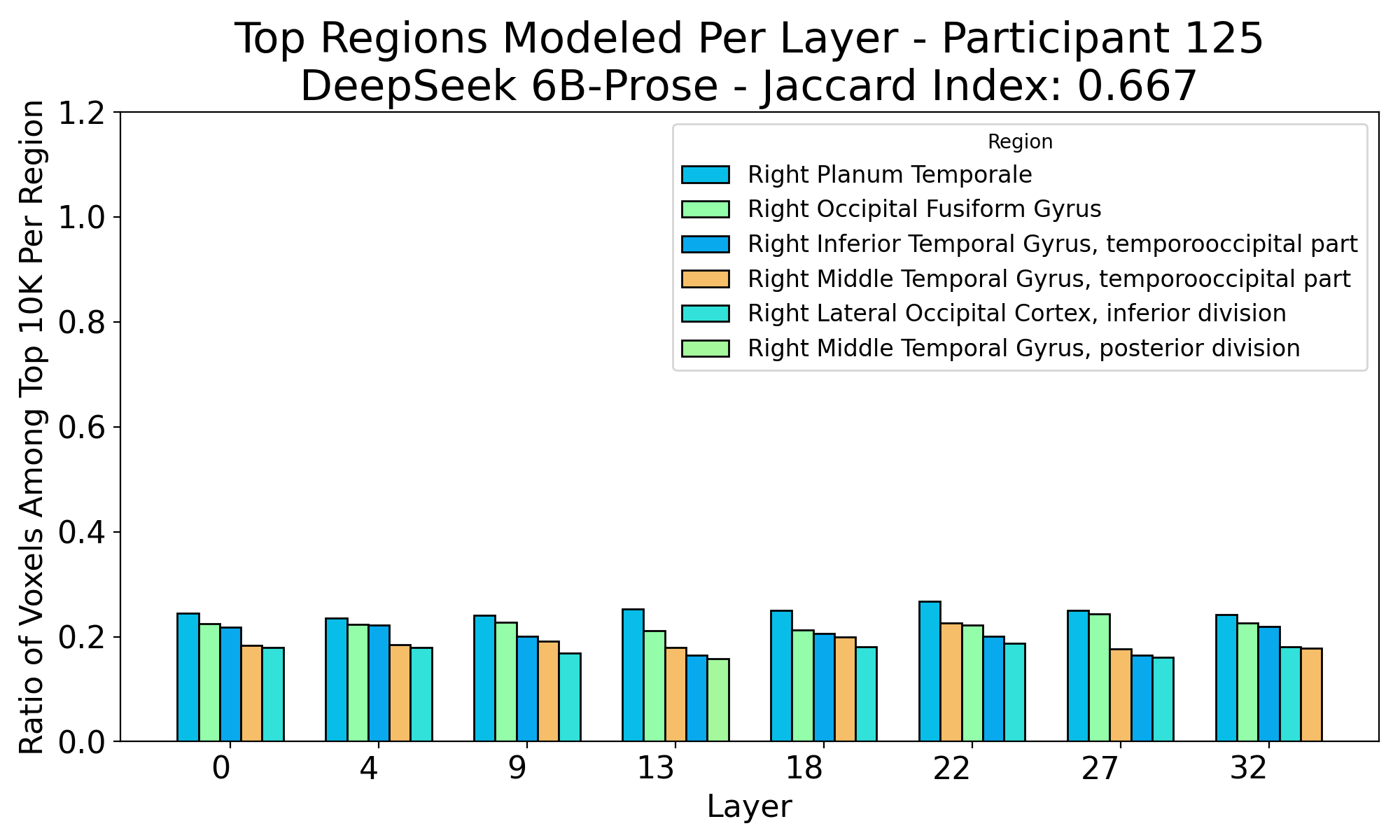}
        \caption{}
    \end{subfigure}
    \caption{Locations of the top 10K modeled voxels in the Harvard-Oxford Atlas, filtered to the top five regions for each layer for the best performing model and parameter configurations. Depicted here is Participant 125 for (a) Code and (b) Prose, whose z-transformed correlation coefficients were lowest for Code among all the participants.}
    \label{fig:bottom_bar_proportions}
\end{figure*}

We also investigated whether LLM embeddings from layers throughout the LLMs were better able to model activity in different brain regions, using the same methods as we did in Section~\ref{rq:layers}.
For high-level interpretability, we computed top-modeled regions with respect to the Harvard-Oxford Atlas, where we mapped participants' top 10K voxels to regions in the atlas. 
For each region, we divided the number of top 10K voxels within that region by the total number of voxels therein.
These findings can demonstrate whether LLM embeddings from different layers may model more specialized cognitive processes during code and prose writing.
From Figures~\ref{fig:top_bar_proportions} and~\ref{fig:bottom_bar_proportions}, we see consistency in the best-modeled regions, but also increased variability compared to results using raw counts.
Considering that the largest quantity of top 10K voxels is located in the right frontal pole, which is involved in high-level planning, normalized results may provide insights into more fine-grained cognitive processes that can be modeled using LLM-embeddings.

Whereas the average Jaccard indices for raw voxel counts were above 0.6 (60\%) for voxel counts in Section~\ref{rq:layers}, the average here for Code is 0.405 ($\sigma=0.314$) and for Prose is 0.440 ($\sigma=0.296$).
In other words, the brain regions that are modeled best by various LLM layers are only about 40\% consistent between the layers.
The differences in Jaccard Indices for Code and Prose did not reach statistical significance ($t=0.381$, $p = 0.707$).
Moreover, we did not find a significant correlation between participants' Jaccard Indices and modeling performance, measured as the average of a participant's top 10K z-transformed correlation coefficients, for Code ($r=-0.033$, $p=0.883$) or Prose ($r=0.090$, $p=0.681$). 
These findings are similar to those above in Section~\ref{rq:layers}.
However, the increased variability suggests that there may be more nuance in what cognitive processes are aligned between brain activity and LLM embeddings in each layer.

\begin{figure*}[htbp]
    \centering
    \includegraphics[width=\textwidth]{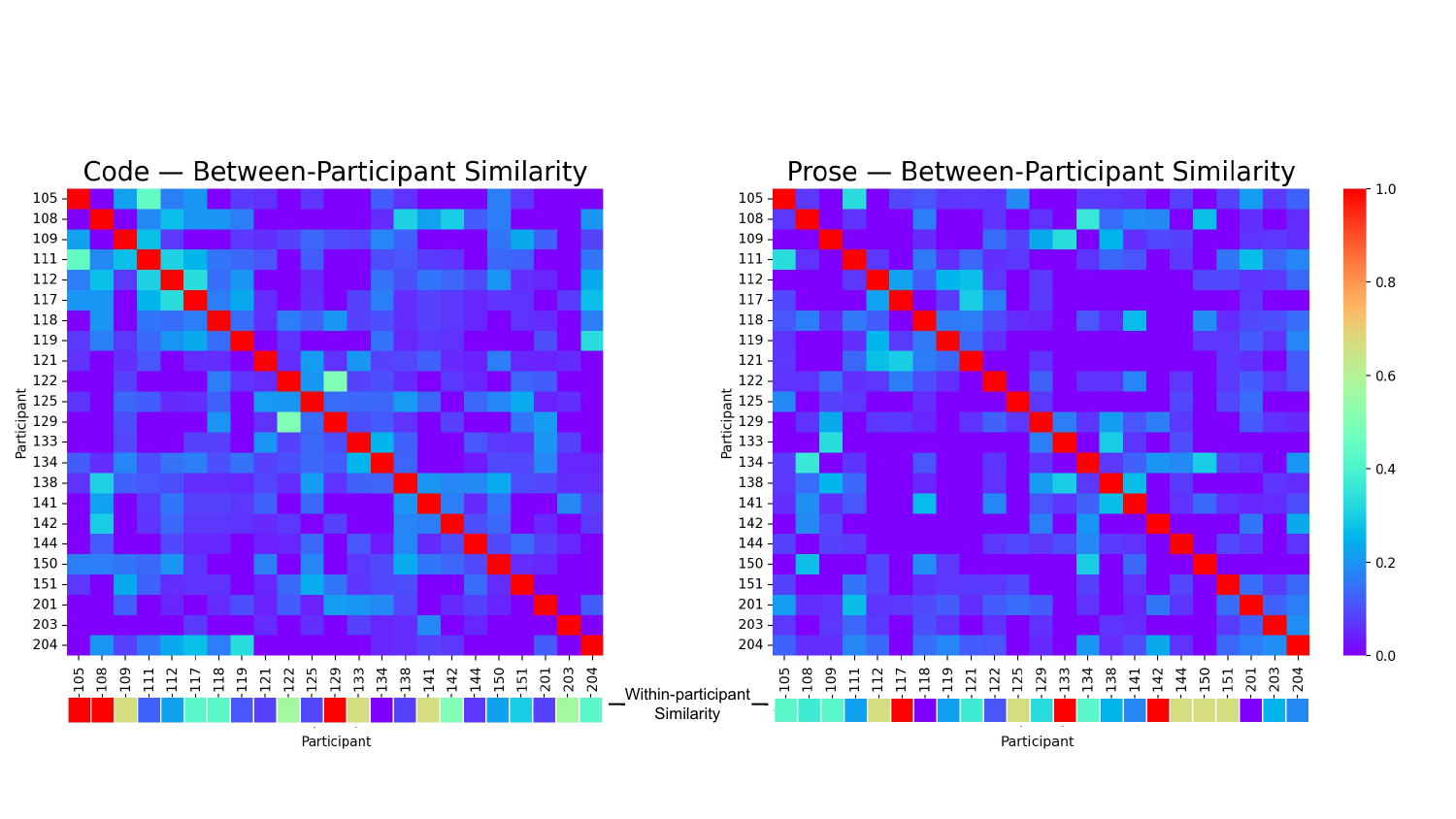}
    \caption{Pairwise similarities in the top regions between all participants. These values were calculated by computing the Jaccard Index between two given participants' top-modeled regions. These sets of regions for each participant were computed by taking the union of their top-modeled regions across the different LLM layers. Also depicted is the within-participant similarity, which is the Jaccard Index across the different model layers for each participant.}
    \label{fig:similarity_proportions}
\end{figure*}

We also computed the between-participant similarity in top-modeled regions, which can be seen in Figure~\ref{fig:similarity_proportions}.
The average similarities here are markedly lower than those using voxel counts. 
This is true for both Code ($\mu = 0.088$, $\sigma = 0.087$) and Prose ($\mu = 0.070$, $\sigma = 0.080$).
Differences here between Code and Prose are statistically significant, based on a paired t-test ($t=2.589$, $p=0.010$), suggesting that between-participant similarity is higher with respect to top-modeled brain regions for code writing compared to prose writing. 
We see that the overall findings echo those above in Section~\ref{rq:layers} about similarity within and between participants. 
Nonetheless, there is higher variability in best-modeled regions both within and between participants when we normalize by region size.
This suggests that top-modeled regions can be more distinctive for a given participant or LLM layer, and using normalized parcels may help uncover meaningful individual differences.

Finally, we computed key metrics for other LLMs we considered in our study, as shown in Table~\ref{tab:other_llms_proportions}.
Here we see similar trends as above, where within-participant similarity ($\mu=0.325$, $\sigma=0.124$) is higher than between-participant similarity ($\mu=0.088$, $\sigma=0.013$), which we found was statistically significant using a paired t-test ($t=4.258$, $p<0.01$, $d=1.739$).
We also observe consistency in the top-modeled regions depending on the task, which contain regions we might expect based previous research, compared to top regions we report above in Table~\ref{tab:other_llms}.
Specifically, for Prose we see that the left middle temporal gyrus and left inferior frontal gyrus are among the top-modeled regions for each model, which are key regions for syntactic and semantic processing~\citep{friederici2013language}.

Moreover, for Code we see that the left superior parietal lobule is consistently the best-modeled region using LLM embeddings, followed by other parietal and temporal regions, as well as the left inferior frontal gyrus.
These parietal and frontal regions have been strongly implicated in previous research for code reading and writing~\citep{siegmund2014understanding, krueger2020neurological, liu2025rapid}. 
These aggregated results suggest that at the group level, top-modeled regions align with regions of interest reported in previous research.
These findings indicate that raw voxel counts may provide insights into general cognitive processes that are aligned with LLM embeddings, but normalized voxel proportions may uncover meaningful contrasts between tasks.

\begin{table}[htbp]
    \small
    \centering
    \caption{Within- and between-participant similarities based on Jaccard Index for other top-performing parameter configurations for Code and Prose, as well as the top-modeled regions for the other LLMs. Results align with our findings for DeepSeek 6B, with the exception of CodeGemma 7B.}
    \label{tab:other_llms_proportions}
    \resizebox{\textwidth}{!}{
    \begin{tabular}{l|l|r|r|r|r|p{8cm}}
        \hline
        \makecell[l]{\textbf{Task}} & \makecell[l]{\textbf{Model}} & \makecell[r]{\textbf{Number of}\\ \textbf{Delayed Copies}} & \makecell[r]{\textbf{Look-Ahead}\\\textbf{Volumes}} & \makecell[r]{\textbf{Within-Participant}\\ \textbf{Similarity}} & \makecell[r]{\textbf{Between-Participant}\\ \textbf{Similarity}} & \makecell[l]{\textbf{Top Modeled Regions}} \\
        \hline
        \multirow{5}{*}{Code} & \multirow{5}{*}{DeepSeek 6B} & \multirow{5}{*}{10} & \multirow{5}{*}{0} & \multirow{5}{*}{$0.405\pm0.314$} & \multirow{5}{*}{$0.088\pm0.087$} & \begin{enumerate}[noitemsep, topsep=7pt, parsep=0pt, partopsep=0pt, leftmargin=*, before=\vspace{-\baselineskip}]  \item Left Superior Parietal Lobule
          \item Left Supramarginal Gyrus, anterior division
          \item Left Inferior Frontal Gyrus, pars triangularis
          \item Right Middle Temporal Gyrus, anterior division
          \item Left Postcentral Gyrus
        \end{enumerate}\\\hline
        \multirow{5}{*}{Code} & \multirow{5}{*}{CodeGemma 7B} & \multirow{5}{*}{10} & \multirow{5}{*}{1} & \multirow{5}{*}{$0.157\pm0.179$} & \multirow{5}{*}{$0.108\pm0.075$} & \begin{enumerate}[noitemsep, topsep=7pt, parsep=0pt, partopsep=0pt, leftmargin=*, before=\vspace{-\baselineskip}]  \item Left Superior Parietal Lobule
          \item Right Occipital Pole
          \item Right Middle Temporal Gyrus, anterior division
          \item Right Lateral Occipital Cortex, inferior division
          \item Right Frontal Medial Cortex
        \end{enumerate}\\\hline
        \multirow{5}{*}{Code} & \multirow{5}{*}{StarCoder 3B} & \multirow{5}{*}{10} & \multirow{5}{*}{0} & \multirow{5}{*}{$0.325\pm0.226$} & \multirow{5}{*}{$0.083\pm0.080$} & \begin{enumerate}[noitemsep, topsep=7pt, parsep=0pt, partopsep=0pt, leftmargin=*, before=\vspace{-\baselineskip}]  \item Left Superior Parietal Lobule
          \item Left Supramarginal Gyrus, anterior division
          \item Left Inferior Frontal Gyrus, pars opercularis
          \item Left Inferior Temporal Gyrus, anterior division
          \item Right Lateral Occipital Cortex, inferior division
        \end{enumerate}\\\hline
        \multirow{5}{*}{Prose} & \multirow{5}{*}{StarCoder 7B} & \multirow{5}{*}{16} & \multirow{5}{*}{3} & \multirow{5}{*}{$0.351\pm0.257$} & \multirow{5}{*}{$0.080\pm0.086$} & \begin{enumerate}[noitemsep, topsep=7pt, parsep=0pt, partopsep=0pt, leftmargin=*, before=\vspace{-\baselineskip}]  \item Left Inferior Frontal Gyrus, pars triangularis
          \item Right Frontal Medial Cortex
          \item Right Occipital Pole
          \item Left Middle Temporal Gyrus, anterior division
          \item Left Frontal Pole
        \end{enumerate}\\\hline
        \multirow{5}{*}{Prose} & \multirow{5}{*}{CodeGemma 7B} & \multirow{5}{*}{16} & \multirow{5}{*}{3} & \multirow{5}{*}{$0.216\pm0.174$} & \multirow{5}{*}{$0.098\pm0.082$} & \begin{enumerate}[noitemsep, topsep=7pt, parsep=0pt, partopsep=0pt, leftmargin=*, before=\vspace{-\baselineskip}]  \item Left Middle Temporal Gyrus, anterior division
          \item Right Frontal Medial Cortex
          \item Left Frontal Medial Cortex
          \item Right Occipital Pole
          \item Left Inferior Frontal Gyrus, pars triangularis
        \end{enumerate}\\\hline
        \multirow{5}{*}{Prose} & \multirow{5}{*}{DeepSeek 6B} & \multirow{5}{*}{16} & \multirow{5}{*}{3} & \multirow{5}{*}{$0.496\pm0.316$} & \multirow{5}{*}{$0.071\pm0.090$} & \begin{enumerate}[noitemsep, topsep=7pt, parsep=0pt, partopsep=0pt, leftmargin=*, before=\vspace{-\baselineskip}]  \item Left Middle Temporal Gyrus, anterior division
          \item Right Frontal Medial Cortex
          \item Right Inferior Frontal Gyrus, pars triangularis
          \item Right Occipital Pole
          \item Right Angular Gyrus
        \end{enumerate}\\\hline
    \end{tabular}
    }
\end{table}

\printbibliography



\end{document}